\documentclass[
twocolumn,
tightenlines,
10pt,
longbibliography,
showpacs,
nofootinbib,
notitlepage,
superscriptaddress]{revtex4-2}
\usepackage{algorithm2e}
\usepackage{colortbl, longtable}
\usepackage{amsmath, amssymb, amsthm, braket, enumitem, dsfont}
\usepackage{makecell, graphicx, subfiles, multirow}
\usepackage{hyperref}
\usepackage[capitalize]{cleveref}

\newcommand{\CNOT}{{\sc CNot}\ }
\newcommand{\SWAP}{{\sc Swap}\ }

\begin{document}

\title{Nuwa: A Quantum Circuit Transpiler
Based on a Finite-Horizon Heuristic for Placement and Routing}

\author{Shengru~Ren}
\thanks{These authors contributed equally to this work.}
\affiliation{1QB Information Technologies (1QBit), Vancouver, BC, Canada}

\author{KaWai~Chen}
\thanks{These authors contributed equally to this work.}
\affiliation{1QB Information Technologies (1QBit), Vancouver, BC, Canada}

\author{Navid~Ghadermarzy}
\thanks{These authors contributed equally to this work.}
\affiliation{1QB Information Technologies (1QBit), Vancouver, BC, Canada}

\author{Brandon~Nguyen}
\affiliation{1QB Information Technologies (1QBit), Vancouver, BC, Canada}

\author{Yanhao~Huang}
\affiliation{1QB Information Technologies (1QBit), Vancouver, BC, Canada}

\author{Pooya~Ronagh}
\email[Corresponding author: ]{pooya.ronagh@1qbit.com}
\affiliation{1QB Information Technologies (1QBit), Vancouver, BC, Canada}
\affiliation{Institute for Quantum Computing, University of Waterloo, Waterloo, ON, Canada}
\affiliation{Department of Physics \& Astronomy, University of Waterloo, Waterloo, ON, Canada}
\affiliation{Perimeter Institute for Theoretical Physics, Waterloo, ON, Canada}

\begin{abstract}
We introduce a novel transpiler for the placement and routing of quantum
circuits on arbitrary target hardware architectures. We use finite-horizon,
and optionally discounted, reward functions to heuristically find a suitable
placement and routing policy. We employ a finite lookahead to refine the reward
functions when breaking a tie between multiple policies. We benchmark our
transpiler against multiple alternative solutions and on various test sets of quantum algorithms to demonstrate the benefits of our approach.
\end{abstract}

\maketitle

\raggedbottom

\section{Introduction}
\label{sec:intro}

Classical optimization plays an important role in the study and development of
quantum computing software--hardware stacks. Various layers of compilers must be used to
transform a quantum algorithm written in a quantum programming language
\cite{larose2019overview} to form an executable algorithm on a quantum processor. Ideally,
this process can be fully automated so that a quantum software
developer can abstract her attention away from considerations of optimal
commutations of quantum gates~\cite{itoko2019quantum, itoko2020optimization},
mitigating various sources of noise and errors on the executed instruction sets~\cite{murali2019noise}, and the specific architectures of backend quantum
devices~\cite{leymann2020bitter}.

An optimization problem related to compiling for a specific architecture requires the
construction of an efficient transformation of the quantum program into one executable
on specific systems of physical or logical qubits with geometric locality
constraints such as 2D or 3D nearest-neighbour connectivity or other limited
sets of multi-qubit entangling gates. This compilation step is of
interest for the near-term compilation of small- to moderate-sized quantum algorithms
on noisy, intermediate-scale quantum (NISQ) systems that have between tens and thousands of
qubits~\cite{preskill2018quantum}. Moreover, similar compilation procedures are of interest for fault-tolerant quantum computation (FTQC) on the
large-scale architectures anticipated for quantum computers, such as modular architectures
comprising arrays of superconducting qubits~\cite{lao2018mapping} or quantum dots and donors~\cite{vandersypen2017interfacing}, and modular trapped-ion
architectures dependent on shuttling ions between different trapping zones~\cite{brown2016co}.

The goal of the aforementioned optimization task is to compile an input quantum circuit,
written as an abstract sequence of quantum gates, into another circuit
comprising quantum gates physically available on the target quantum processor
while incurring the least possible overhead. A compiler that performs this translation may instead be called a
\emph{transpiler}, especially when the circuit transformation is happening at the
same level of software abstraction (e.g., when both the input quantum circuit and the
compiled circuit are written in terms of the \mbox{Clifford + $T$} gates), as opposed to
when the compilation is from a higher level of abstraction to a lower one. The
transpiler may aim to reduce the total number of additional gates it introduces
in order to minimize the number of erroneous operations performed. Alternatively,
the transpiler may reduce the depth, or \emph{makespan},
of the compiled circuit in order to reduce the overall runtime of the algorithm,
assuming that simultaneous gates can be performed with a negligible amount of cross-talk.
In this paper, we focus on the former optimization objective of reducing the total number of additional gates the transpiler introduces.

This optimization problem is often modelled as a \emph{placement and routing}
problem of finding a mapping of the abstract qubits to the
physical qubits (placement), followed by iterations of performing entangling
gates between qubits that are far apart by moving them closer to each other (routing),
for example, via \SWAP gates. The placement and routing problem has been extensively
studied in the context of NISQ algorithms~\cite{shafaei2014qubit,
wille2014optimal, farghadan2017quantum, siraichi2018qubit, lao2018mapping,
herbert2018using, zulehner2018efficient, booth2018comparing,
cowtan2019qubit,murali2019noise, childs2019circuit, venturelli2019quantum,
li2019tackling, niu2020hardware, itoko2020optimization, paler2021nisq,
sinha2021qubit, lao20222qan, molavi2022qubit, nannicini2022optimal,
wagner2022improving, zhou2020monte}, as well as in the fault-tolerant
compilation of quantum algorithms using concatenated error-correcting
codes~\cite{lin2014paqcs}, lattice surgery--based fault-tolerant quantum
\mbox{computing (FTQC)~\cite{paler2017online, lao2018mapping},} and defect-based
FTQC~\cite{javadi2017optimized}.

Additionally, the placement problem has been specifically studied in
\cite{gerard2021string, tan2021optimal, gerard2021exploring, fan2022optimizing}.
Our focus in this paper is on developing techniques for routing, although we
use them to choose good placements as well as in cases where multiple options are
available. We should also distinguish the  literature cited herein and our
paper from the works \cite{wu2020qgo, weiden2022wide, xu2022synthesizing,
van2022dynamic, gheorghiu2022reducing}, wherein architectural considerations
are incorporated with compilers operating at the circuit synthesis level.

The problem and several of its variants  are known to be
NP-hard~\cite{siraichi2018qubit, cowtan2019qubit}. Therefore, several of the mentioned
references rely on heuristic methods for finding a \mbox{(sub-)optimal} placement and
routing policy. Some of the solutions relevant to our approach
are those based on the observation that the placement and routing problem is an
inherently temporal problem over the span of multiple decision epochs. This
includes an exact, but exponentially expensive, dynamic programming solution
introduced in~\cite{siraichi2018qubit}, temporal planning and constraint
programming~\cite{booth2018comparing, venturelli2019quantum}, integer
programming~\cite{nannicini2022optimal}, and methods
that use reinforcement learning~\cite{herbert2018using, sinha2021qubit}. On one
hand, these approaches are superior to the greedy techniques that neglect the
future impact of decisions made at earlier decision epochs. On the other hand, the
global nature of the optimization problem solved hinders the scalability of the
algorithm, especially in instances where the compilation procedure is expected to
include computationally intensive subroutines such as the training of neural
networks.

In this paper, we propose a middle-ground solution to the problem that captures
both the efficiency of heuristic and greedy approaches, and the optimality of
dynamic programming solutions in temporal decision making problems. This
results in an algorithm that scales favourably in the \SWAP gate
overhead compared to other algorithms. As shown in \cref{sec:results} and
\cref{sec:results-tables}, our algorithm achieves superior scaling compared to
the state-of-the-art traspilers for general quantum circuits
\cite{childs2019circuit, sivarajah2020t}. We note that while
\cite{zhou2020monte} and \cite{molavi2022qubit} show improved results over those of
\cite{sivarajah2020t} for small circuits, a scalable performance improvement is
not evident from these papers. In addition, \cite{lao20222qan}
and 
\cite{wagner2022improving} customize the transpilation problem for 2-local
Hamiltonian simulation circuits whereas our algorithm is not specific to the
type of input quantum algorithm (see also \cref{compare_childs_pytket_nuwa}
for the benchmarking results of our algorithm on different types of quantum circuits).

\subsection{An Overview of Our Algorithm}

Although our algorithm is a heuristic search method, it is inspired by
techniques in approximate dynamic programming~\cite{powell2009you},
as will be apparent from the terminology we use. We introduce two \emph{reward
functions}, the placement score~\cref{eq:placement}
and edge score~\cref{eq:edge-score}, that
resemble the (possibly discounted) cummulative reward structures in dynamic
programming and reinforcement learning. However, instead of performing costly
exact or approximate dynamic programming recursions on them, we restrict their
definitions to a finite \emph{decision horizon} in which only a finite number
of subsequent two-qubit gates are taken into account. We then use
heuristic search to decide on the immediate actions to take, that is, the
placement and routing of a subsequent \emph{layer} of gates. The full placement
and routing \emph{policy} is formed by iterating over this process.
The reward functions introduce two hyperparameters, namely, a
\emph{discount factor} and the decision horizon. These hyperparameters are used to tune
the significance of future gates within the decision horizon on the values of the
reward functions and, consequently, on the routing policy.

Calculating the finite-horizon
reward functions frequently results in a tie across multiple routing options. This result
is more prominent in the case of more-structured quantum circuits, which constitute most of
the practical input circuits of the transpiler. Our algorithm uses a tie-breaking subroutine (\cref{sec:tiebreak}) to refine the routing decisions. We use a hyperparameter called the \emph{lookahead depth} to determine how many decisions to take into account (i.e., the number of \SWAP gates we intend to insert) in order to refine the evaluation of our immediate actions when a tie occurs between the reward function values.

Our transpiler is agnostic with respect to hardware architectures; it is capable of receiving a representation of
the connectivity of any target architecture as an input. Moreover, the
hyperparameters decision horizon, discount factor, and lookahead depth
allow our algorithm to adapt to the structure and recurrent patterns of the
input quantum circuits and the classical computing budget
available to it in order to provide a suitable (sub-)optimal placement and
routing policy.

\subsection{Related Literature}

We provide a detailed description of our approach in~\cref{sec:main}, and
in~\cref{sec:results} we demonstrate the superior performance of our
transpiler against other algorithms proposed in the recent literature that, to
our knowledge, are representative of the state of the art. Having said that,
our approach incorporates several ideas from these competitors, which in this section we summarize comparatively.

We use two distinct routines for placement and routing. Our
algorithm first chooses a placement by generating a set of candidates and ranking them according to their placement scores (\cref{sec:placement}).
In principle, any placement can be considered in this ranking, but in this paper we confine our routine to generating candidate placements using the approaches of
\cite{cowtan2019qubit} and \cite{childs2019circuit}. \cite{cowtan2019qubit} uses a linear path built from the connectivity between qubits to find the
initial placement, and in \cite{childs2019circuit} a matching-based method is used (see \cref{sec:linear_placement} and \cref{sec:matching_placement} for
more-detailed descriptions of the two placement methods). Neither
of these references provide or utilize a systematic metric for comparing
multiple placements with each other, whereas our placement score incorporates the effect of future gates in the input circuit for the purpose of this comparison.

In contrast, \cite{li2019tackling} iterates between a forward and backward
traversal of the sequence of two-qubit gates in order to generate a placement
that takes future gates into account. This procedure is used to update the
initial placement while routing the qubits in the forward and backward passes
using a heuristic. As such, \cite{li2019tackling} does not use separate stages
for placement and routing; instead, the two tasks are intertwined. Iterating
over this process becomes more expensive as the depth of the circuit grows and
there is no garauntee that a greater number of iterations will produce better transpilations of the
input circuit. On the other hand, \cite{li2019tackling} uses two sets of future
gates (in either the forward or reverse passes), the gates to be processed immediately in the
execution order and a subsequent set of future gates, to rank possible routing
options. The idea of using future gates in a heuristic search to make better
routing decisions is systematically incorporated in our work using the
(discounted and cummulative) edge scores with a tunable hyperparameter that
varies the impact of gates in future layers (see \cref{sec:device} and \cref{sec:routing} for the
definitions of the concepts of layers and edge
scores, respectively).

Unlike the strategy taken in \cite{li2019tackling}, \cite{childs2019circuit}
presents the finding that a routing scheme chosen in a greedy fashion performs better overall in numerical benchmarking. In \cite{cowtan2019qubit}, the routing policy is also chosen greedily and based on the two-qubit gates to be processed immediately in the execution
order. A distance vector is incorporated in this process. The \SWAP
gate is chosen from a set of candidates that reduces the greatest distance
between pairs of qubits that are invovled in two-qubit gates in line to be executed.
The process is continued until either there is only one candidate left or a
predefined limit has been reached. If this method fails, in \cite{cowtan2019qubit}
the same strategy is repeated on pairs of \SWAP gates instead of individual ones,
and if this also fails it resorts to brute force by routing maximally distant
qubits involved in a two-qubit gate towards each other in order to escape this situation.

We also use a similar myopic strategy for routing (see the definition of
immediate edge scores in \cref{sec:routing}). However, according to our findings, symmetries
of the input circuit can frequently cause degenerate decisions in a myopic routing strategy. This is why in our tie-breaking subroutine
(\cref{sec:tiebreak}) we use the more general edge scores (with discounted
contributions from gates in the further layers included) to choose the edge
that leads to the minimum number of future \SWAP gates needed within the
decision horizon.

Finally, it is worth noting that \cite{li2019tackling} and
\cite{childs2019circuit} consider the trade-off between optimizing the
number of inserted \SWAP gates and the depth of the circuit (assuming
indepedent gates can be executed concurrently). However, in this
paper, we focus on optimizing the total number of additional gates, assuming
the near-term and intermediate-scale scenarios in which the quantum state
can to some extent be protected against decoherence (using techniques such as
dynamic decoupling, error mitigation, and error correction). However, higher
\SWAP overhead and the cross-talk caused by performing multiple gates
concurrently are both more detrimental to the performance of a quantum device.

\section{The Placement and Routing Problem}
\label{sec:device}

\begin{figure}[b]
  \centering
  \includegraphics[trim=3cm 17cm 2cm 5cm, clip, width=\linewidth]{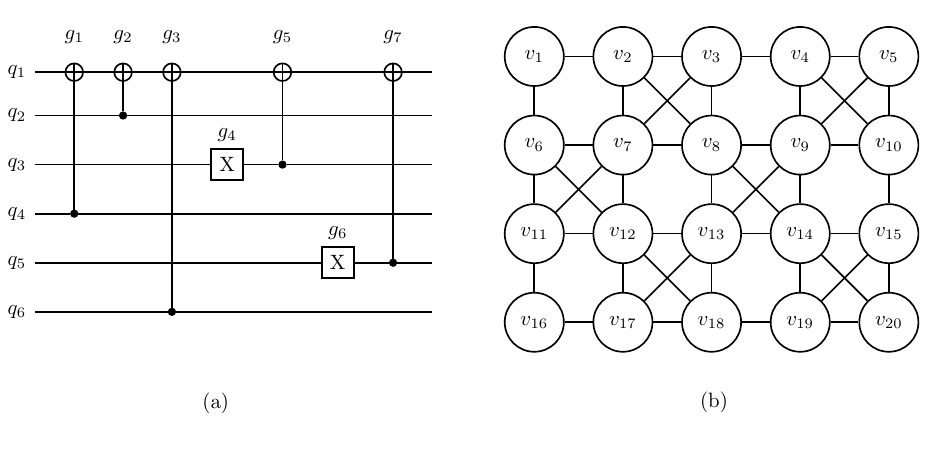}
  \vspace{-1cm}
  \caption{(a) Example circuit with seven gates and six qubits.
  (b) Undirected coupling graph of the IBM Q20 Tokyo chip.}
  \label{fig:graph_and_circuit}
\end{figure}

The problem of interest in this paper is as follows. Given an input quantum
circuit, or \emph{input circuit}, we wish to find a second
circuit, or \emph{compiled circuit}, that executes the  quantum algorithm but
uses the quantum gates available on a physical device, called the \emph{target
architecture}, or architecture for short. We assume that the input circuit is
determined by a sequence
\begin{equation}
\sigma= (g_1, \ldots, g_N)
\end{equation}
of $N$ single-qubit rotations and \CNOT gates afflicting a set
\begin{equation}
Q= \{q_1, \ldots, q_M\}
\end{equation}
of $M$ (logical) qubits. The target architecture of the
physical device also allows it to perform similar gates, but \CNOT gates are not
available between all pairs of physical qubits.
The goal of our algorithm is to generate
a compiled circuit that can be executed on the target architecture by
inserting a plurality of \SWAP gates in the sequence.
We note that every \SWAP gate can be performed
using three \CNOT gates. As the addition of gates creates a deeper circuit, the
compiled circuit is more prone to errors and takes a longer time to execute.
Therefore, our compiler has to minimize the number of inserted \SWAP gates, from
here on called the \emph{\SWAP count}, $N_\text{S}$.

Figure~\ref{fig:graph_and_circuit}(a) shows a representation of an input circuit.
The target architecture to which
we intend to compile our quantum algorithm is represented using an undirected
graph called a \emph{coupling graph}, $G= (V, E)$, with vertices $V$ and
edges $E$. Each physical qubit is represented as a
vertex $v \in V$ and the two-qubit gates available between physical
qubits are represented by the edges $\{u, v\} \in E$ between the vertices $u$ and $v$. We
do not consider direction for the edges in this paper; however, our algorithm can be
adapted to the case of devices with a particular direction for two-qubit gates
as in the case of~\cite{dueck2018optimization}. Figure~\ref{fig:graph_and_circuit}(b)
shows the coupling graph of the IBM Q20 Tokyo chip as an example target architecture~\cite{hillmich2021exploiting}.

\begin{figure}[t]
  \centering
  \includegraphics[trim=3.5cm 18.5cm 4cm 4.5cm, clip, width=\linewidth]{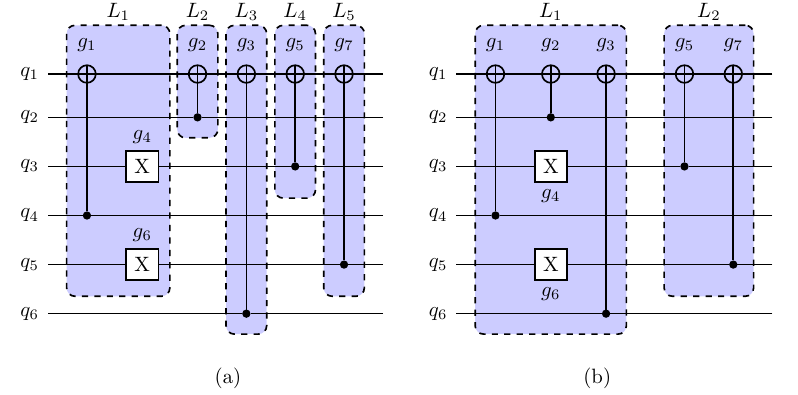}
  \vspace{-0.5cm}
  \caption{Example circuit after gates are grouped into layers using
  (a) fine layering and (b) coarse layering.
  Each highlighted
  box represents a layer. Gates in a given highlighted box belong to the same layer.}
  \label{fig:layer_example}
\end{figure}

\begin{figure}[b]
  \centering
  \includegraphics[trim=4.5cm 19.5cm 4cm 4.5cm, clip, width=\linewidth]{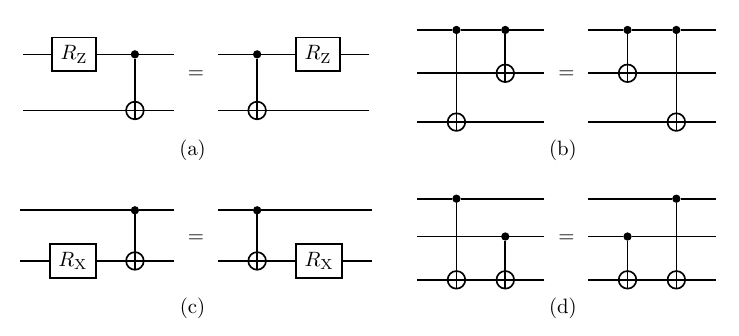}
  \vspace{-0.5cm}
  \caption{(a) $R_\text{Z}$--control,
  (b) Control--control,
  (c) $R_\text{X}$--target, and
  (d) Target--target.}
  \label{fig:comm_rules}
\end{figure}

We construct a \emph{layering} for the input circuit, which is a partition
\begin{equation}
L= \bigsqcup_{i=1}^{N_L} L_i
\end{equation}
of the set of all gates in the input circuit $\sigma$ into $N_L$ layers.
Each layer comprises a subset of the gates in $\sigma$. The layering
is alternatively given by a function
\begin{equation}
\ell\!: \{g_1, \ldots, g_N\} \to \mathbb N,
\end{equation}
where $\ell(g_i)$ is the index of the layer to which gate $g_i$ belongs. We perform
this decomposition in two ways. In both cases, we assure the minimal number of
layers are constructed. The construction also guarantees that the gates in the
same layer can be executed in any order.

In the first decomposition, called \emph{fine layering}, gates that share
qubits are grouped into distinct layers. Figure~\ref{fig:layer_example}(a) shows an
example of fine layering. The single-qubit gates are moved to the earliest layer to which
they can be moved. The single-qubit gates do not affect the placement and routing procedure explained in this paper.

In the second decomposition, called \emph{coarse layering}, gates that
mutually commute are allowed to be in the same layer. The commuting
relations we use are the same ones used in~\cite{itoko2020optimization}, and are
shown in~\cref{fig:comm_rules}. An example of coarse layering can be seen in \cref{fig:layer_example}(b). The motivation for using coarse layering is that all
commuting gates within the same layer can be executed in any order, thus
reducing the potential \SWAP count, $N_\text{S}$.

\section{Description of Our Algorithm}
\label{sec:main}

Our algorithm consists of two main
subroutines: placement and routing. We describe our method for finding a good placement
in~\cref{sec:placement}, and then present our routing subroutine in~\cref{sec:routing}.

\subsection{Placement}
\label{sec:placement}

The first step of our algorithm is to assign each qubit in the circuit
to a vertex. An injective mapping
$$p\!: Q \to V$$
of qubits involved in the input circuit $\sigma$ to vertices of the coupling graph is
called a \emph{placement}. To a placement $p$ we associate a \emph{placement
score}
\begin{equation}
\label{eq:placement}
\Sigma_{p, \ell}(\lambda, \delta)=
\sum_{q \in Q}
\sum_{g \in \sigma_{q, \lambda}} \delta^{\ell(g) - 1}
d_G\left(p(g_1), p(g_2)\right),
\end{equation}
where $\lambda$ is a positive integer called the \emph{decision horizon}, or
\emph{horizon} for short. For each qubit in $Q$, the horizon represents the
number of future gates in the circuit taken into account in order to evaluate
the utility of a placement. For every qubit $q \in Q$, the set
$\sigma_{q, \lambda} \subseteq \sigma$ is the subset of the first $\lambda$
gates $g$ in $\sigma$ acting on $q$ with a unique other qubit involved. Here,
$g=(g_1, g_2)$ is a gate acting on qubits $g_1$ and $g_2$, and $d_G$ denotes
the distance between vertices in the coupling graph $G$. The value
$\delta \in [0, 1]$ is called the \emph{discount factor}, and represents the
significance of the contribution of later gates to the placement score.
Discount factors closer to $0$ aggressively suppress the contribution of gates
in future layers, sorting them in chronological order, roughly speaking, whereas
discount factors closer to $1$ assign larger, yet exponentially decaying, weights
on the distances between qubits involved in gates in future layers, allowing the future gates to make greater contributions. The best choice of discount
factors depends on the structure of the input circuit.
The hyperparameters decision horizon and discount factor
appear also in the definition of edge scores in~\cref{sec:routing}.
We study the effect of these hyperparameters on the performance of our algorithm
in \cref{sec:param_selection}.

A good placement is one that can lead to a significant reduction in the final \SWAP gate count. We
present two heuristics for generating placements:~\emph{matching-based}
and~\emph{linear} placement. We
generate a number of placements using the two heuristics. We then compare the
placements with each other using the placement score \cref{eq:placement}.
Afterwards, we use the placement with the lowest placement score to route the circuits.

\subsubsection{Matching-based placement}
\label{sec:matching_placement}

This method involves finding a maximal matching on the coupling graph, and
constructing a placement that maps the gates of the first layer onto the
edges of the matching. The idea of using a matching in the coupling graph was
first introduced in~\cite{childs2019circuit}.

We employ two methods for finding a matching. The first is a greedy approach, in
which edges are added to the matching until no additional edges can be added. The
second approach is to use Edmonds’ blossom algorithm~\cite{edmonds1965paths} to
construct a maximum matching. In general, a maximum matching covers a larger
number of target qubits. However, the blossom algorithm takes more computation time
than the greedy matching and its effect is not significant for highly
structured coupling graphs such as regular lattices. We refer the reader to~\cref{tab:initial_placement_comparison} for a comparison of the performance of
the greedy and blossom matching methods.

Using a matching in the coupling graph might not result in a complete placement.
The remaining qubits without an assigned placement can be mapped to the
remaining vertices at random. However, this may cause the qubits to be assigned
to vertices that are far apart, especially when the number of vertices is much larger
than the number of qubits. This would require us to introduce unnecessary \SWAP
gates in the compiled circuit in order to move the qubits closer to each other.
To prevent this behaviour, we introduce another heuristic to keep the placed
qubits closer to the more-connected regions in the architecture.

\subsubsection{Linear placement}
\label{sec:linear_placement}
\begin{figure}[b]
  \centering
  \includegraphics[trim=3cm 20cm 3cm 4cm, clip, width=\linewidth]{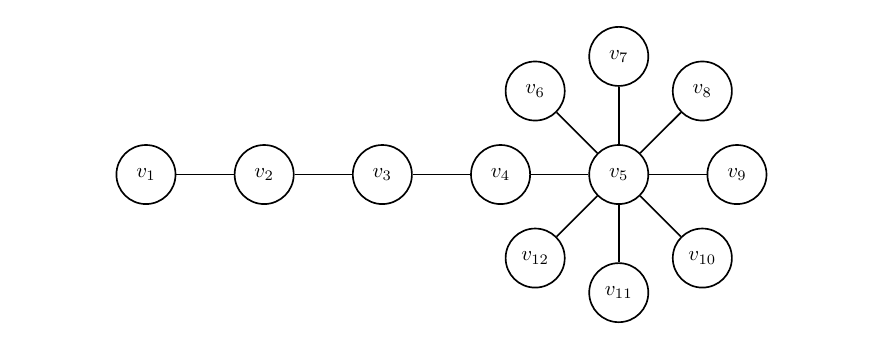}
  \caption{Example architecture containing a star-like subgraph. Starting
  path construction from $v_5$ might result in a short path, whereas starting
  from $v_1$ will result in a longer path.}
  \label{fig:chain}
\end{figure}

We construct maximal paths in the coupling graph using a greedy algorithm. This is done in two
ways: the algorithm greedily favours either the high-degree or low-degree vertices. The latter method is used to account for architectures
that have star-like subgraphs, for which starting from the highest-degree vertices
might result in very short paths (see~\cref{fig:chain} for an example). For more-complex architectures, this greedy approach can be performed continually until a maximal linear forest has been found.

After the paths are constructed, the placement algorithm iterates through the
vertices of any given longest path and moves the qubits to
the earliest free vertices on the path.

As demonstrated in~\cref{sec:placement_comparison}, the two heuristics introduced above can be
used as stand-alone methods or in combination. Optionally, a plurality of
placements can be generated and compared according to their placement scores.

\subsection{Routing}
\label{sec:routing}

The placement provided in~\cref{sec:placement} is used as input for our routing
subroutine. Its pseudocode is shown in~\cref{routing_code}. The goal of
the algorithm is to identify a set of edges in the coupling graph along which
\SWAP gates will be inserted.

\RestyleAlgo{ruled}
\LinesNumbered
\let\oldnl\nl % Store \nl in \oldnl
\newcommand{\nonl}{\renewcommand{\nl}{\let\nl\oldnl}}%
\begin{algorithm}[]
  \SetKwInput{KwData}{Input}
  \SetKwInput{KwResult}{Output}
  \KwData{Input circuit $\sigma$,
  target architecture $G$, a layering $L= \sqcup L_i$, and a placement $p$}
  \KwResult{Compiled circuit $\tau$}
  Initialize the compiled circuit $\tau$ as an empty sequence\;
  \While{$\sigma$ \textup{is not empty}} {
      Construct a sequence $\rho$ of gates in $L_1$ already executable on $G$\;
      Update $\tau$ by appending $\rho$ to it\;
      Update $\sigma$ by removing $\rho$ from it\;
      Recompute the layering for $\sigma$\;
      \For{\textup{every free qubit} $q$ \textup{in the input circuit}}{
        Change the placement of $q$ if a different placement improves
        the placement score\;
      }
      \If{\textup{there are no executable gates in} $L_1$}
      {
          Calculate the edge scores for edges incident to vertices involved in $L_1$; \\
          \uIf{\textup{the optimal edge score is degenerate}}
          {
              Use subroutine presented in \cref{sec:tiebreak} to break the tie\;
          }
          \Else
          {
              \textup{Append a \SWAP gate along the optimal edge to $\tau$}\;
          }
      }
  }
  \caption{\,Routing Algorithm}
  \label{routing_code}
\end{algorithm}

\begin{figure}[b]
  \centering
  \includegraphics[trim=3cm 18cm 2cm 4cm, clip, width=\linewidth]{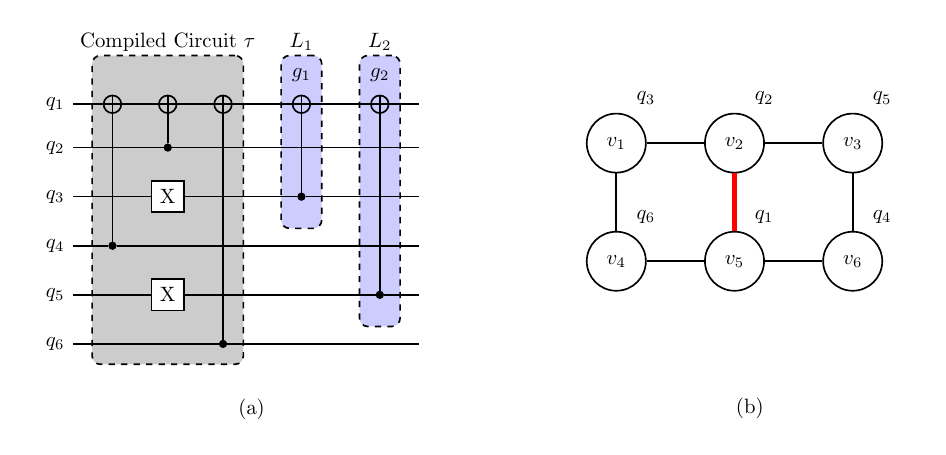}
  \caption{(a) Example circuit with
  gates in $\tau$ already routed,
  gate $g_1$ in layer $L_1$, and gate $g_2$ in layer $L_2$.
  (b) Mapping of the circuit on a $2\times3$ grid.
  We first need to identify the gate(s) in $L_1$, which is $g_1$ in this example.
  To perform an edge score calculation, we inspect
  edges $(v_1,v_2)$ and $(v_1,v_4)$, which are incident to the vertex of $q_3$,
  and edges $(v_4,v_5)$, $(v_2,v_5)$, and $(v_5,v_6)$,
  which are incident to the vertex of $q_1$.
  Here, we show how to calculate the edge score for  edge $(v_2,v_5)$, shown in red.
  We denote the edge by $e_{25}=(v_2,v_5)$.
  For qubit $q_1$, the next not-yet-routed gate
  is gate $g_1$, which is in layer $L_1$, and the other qubit
  in $g_1$ is $q_3$, which is mapped to vertex $v_1$. Using
  \cref{eq:edge-score},
  the immediate edge score is then calculated to be
  \mbox{$\Omega_{p, \ell}(e_{25})= d_G(v_1,v_5) - d_G(v_1,v_2)= 1$.}
  Similarly, for the discounted score, the next not-yet-routed gate
  after $g_1$ is $g_2$, which is in layer $L_2$,
  and the other qubit in the gate is $q_5$, which is mapped to
  vertex $v_3$. The edge score that includes gates in both $L_1$ and $L_2$
  is then calculated to be $\Omega_{p, \ell}(\lambda= |L_1|+|L_2|, \delta=0.1, e_{25}) =
  \left(d_G(v_1,v_5)-d_G(v_1,v_2)\right)
  +0.1 \left(d_G(v_3,v_5)-d_G(v_2,v_3)\right) = 1.1$.
  As for the other vertex $v_2$ of $e_{25}$, given that it does not contain a
  qubit which corresponds to the gate in $L_1$, no calculation is needed.}
  \label{fig:edge_score_eg}
\end{figure}

The routing algorithm receives an input circuit $\sigma$ and iteratively
constructs a compiled circuit $\tau$. The algorithm begins by finding \emph{executable}
gates in the first layer, $L_1$. Here, a gate in $L_1$ is called executable if
the qubits involved in the layer are mapped by the placement function to adjacent
vertices on the coupling graph. If we find executable gates, they are removed
from $L_1$ and added to $\tau$ in the order they appear in $\sigma$.

We then revise the layering, which updates $L_1$. At this point,
if there are more vertices than qubits, some of the vertices will not be in the
image of the placement function. We note that, if a qubit has not yet been
used in any gate in $\tau$, the qubit is free to be reassigned to an
unoccupied vertex or exchange places with another unused qubit. We call such qubits~\emph{free qubits}.

After updating $L_1$, for every free qubit found in the input circuit, we
ascertain whether a different placement for the free qubit improves its placement score,
in which case we update the placement of the qubit. If all qubits in $\sigma$
have already been included in $\tau$, as there are no free qubits, this heuristic step will not be executed.

If no executable gates are found, the algorithm finds a number of edges to
insert \SWAP gates along such that the qubits involved in  non-executable gates are swapped to
adjacent vertices, rendering them executable.
Since \SWAP gates can be inserted on any edge in the coupling graph,
we need to determine which edge would be the best next choice.

An edge $e= \{q_1, q_2\} \in E$ in the coupling graph is a good choice if it
results in fewer future \SWAP gates according to our routing algorithm. We
determine such an edge using its \emph{edge score}, which we define as a
summation of two terms
\begin{equation}
\label{eq:edge-score}
\begin{split}
\Omega_{p, \ell}(\lambda, \delta, e)
= \sum_{i= 1, 2} \Omega_{p, \ell} (\lambda, \delta, e, q_i).
\end{split}
\end{equation}
Here, each $\Omega_{p, \ell}(\lambda, \delta, e, q_i)$ is zero if the first gate
(not necessarily $e$) involving $q_i$ is not in $L_1$, and otherwise
\begin{equation}
\label{eq:edge-qubit-score}
\begin{split}
\Omega_{p, \ell}(\lambda, \delta, e, q_i)
= \sum_{g\in \sigma^r_{q_i, \lambda}}
&\delta^{\ell(g) - 1}
\Big[d_G \left(p(g_1), p(g_2)\right)\\
&- d_G \left(p_\text{e}(g_1), p_\text{e}(g_2)\right)\Big].\\
\end{split}
\end{equation}
Here, $\sigma^r_{q, \lambda} \subseteq \sigma$ is the set of the first $\lambda$
gates in $\sigma$ that act on $q$ and one other qubit. The current placement $p$ and the new placement $p_\text{e}$ are associated with the edge $e$ by
switching the placement of the qubits mapped to $q_1$ and $q_2$, if any exist.
That is,
$p_\text{e} (p^{-1}(q_1))= q_2$ if $q_1$ is in the image of $p$, and, similarly,
$p_\text{e} (p^{-1}(q_2))= q_1$ if $q_2$ is in the image of $p$.

Unlike in the definition of $\sigma_{q, \lambda}$ in \cref{sec:placement}, we
allow the repetition of gates in the definition of $\sigma^r_{q, \lambda}$, as it
is natural to expect that each gate between the same qubits should affect the
edge score. We also empirically verify that not enforcing the condition
$\Omega_{p, \ell}(\lambda, \delta, e, q_i)= 0$ when the first gate involving
$q_i$ is not in $L_1$ results in worse performance for our algorithm.

Figure~\ref{fig:edge_score_eg} provides an example of the edge score calculation.
Here, $\lambda$ represents the decision horizon and $\delta$ represents the
discount factor. The extreme case of $\delta= 0$
is of particular interest, as it corresponds to the ``myopic'' decision horizon in
which only the gates in the first layer are taken into account. The edge scores
\[\Omega_{p, \ell} (e):= \Omega_{p, \ell} (\lambda \geq |L_1|, \delta= 0, e)\] are called \emph{immediate} edge scores.

We use the immediate edge scores to heuristically
find a convenient edge along which to insert
a \SWAP gate. It is sufficient to calculate the edge scores only for the edges that
are incident to the vertices corresponding to the gates in $L_1$. This is
because inserting \SWAP gates along edges that are not incident to qubits in $L_1$ has
no impact on $L_1$. We insert a \SWAP gate along the edge with the maximum immediate edge
score. If
multiple edges are found to have the maximum edge score, we invoke a tie-breaking heuristic,
which we explain in~\cref{sec:tiebreak}.

The routing algorithm iterates over the above procedure until there are no
gates left in $L_1$, at which point all gates have been routed and the
algorithm terminates.

\subsubsection{Tie-breaking subroutine}
\label{sec:tiebreak}

In many cases, multiple edges are found to have the maximum edge score. We call such edges \emph{degenerate}.
To break a tie between degenerate edges, the
algorithm temporarily inserts a \SWAP gate for each such edge. It then continues
the routing procedure, using an edge score with a decision horizon $\lambda > |L_1|$,
until it generates a given number $D>0$ of additional \SWAP gates, or it
terminates when the entire input circuit is compiled. The hyperparameter $D$ is
called the \emph{lookahead depth}. For each degenerate edge, the goal of this multi-step lookahead is to
inspect the future state of the placement score after the insertion of $D$ \SWAP gates within the
decision horizon. We note that, compared to $\lambda$, $D$ has to
be sufficiently small for this $D$-step lookahead to remain within the decision
horizon.

If, after the addition of $D$ \SWAP gates, the entire input circuit has
not been routed, the placement score of the resulting placement is stored as
a \emph{tie-breaking score} for the degenerate edges.

However, if the entire input circuit has been routed with $D'\leq D$
additional \SWAP gates, then the tie-breaking score of the degenerate edge is defined
as \mbox{$D'- D$.} The tie-breaking procedure then selects the edge with the
smallest tie-breaking score. If a degenerate case still occurs for the tie-breaking
scores, then one of the degenerate edges is selected at random.

\subsubsection{Complexity analysis}
The time complexity of the routing procedure depends on the size of the input circuit,
the number of steps required to calculate the insertion of a \SWAP gate, and the number of tie-breaking
subroutine calculations. To estimate the upper bound, we assume the
worst case in our analysis. In this case, all qubits are involved in each
step of the calculation. The routing procedure starts by performing an update on the layering, which has a complexity of $O(\lambda N)$, where $N$ is the number of qubits.
The computation of the immediate edge score  has a complexity of $O(N)$. Assuming all edge scores are
degenerate, we call the tie-breaking subroutine for all candidate edges. The subroutine's complexity depends
on the lookahead depth $D$. For each of the candidate edges,  we need to insert at most $D$ \SWAP gates.
This leads to a time complexity of $O(DN)$. In total, the time complexity
for routing a single gate is $O(\lambda N) + O(N) + O(DN)$. Using $N_\text{G}$ to denote the total number of gates, the time complexity of our routing procedure is $O(N_\text{G}\cdot(\lambda + D)N)$.

\section{Numerical results}
\label{sec:results}

In this section, we present our results and compare them with those of other placement
and routing algorithms. To the best of our knowledge, SABRE~\cite
{li2019tackling} is a commonly used quantum compiler for benchmarking , \texttt{arct}~\cite
{childs2019circuit} scales well, and runs fast for large circuits, and \texttt
{tket}~\cite{sivarajah2020t} is the state-of-the-art library for general
quantum circuits.

In~\cref{sec:placement_comparison}, we discuss the effects of different combinations of the
placement schemes,  introduced in~\cref{sec:placement}, on the \SWAP count. As part of an
ablation study, we benchmark the performance of our algorithm both with and
without the application of the commutation rules presented
in~\cref{fig:comm_rules}. Then, in~\cref{sec:main-results}, we use a set of hyperparameters
that provides a good balance between performance and runtime, and the best
placement scheme found to compare our algorithm
against the above-mentioned alternatives. We perform a hyperparameter study
and find that while certain choices of hyperparameters are evidently inadequate
(e.g., the discount factors $\delta= 0$ or $1$), small circuits are not very
sensitive to the values of the hyperparameters. The results of the study can be
found in~\cref{sec:param_selection}.

\subsection{Placement Heuristics}
\label{sec:placement_comparison}

In our experiments, we benchmark three placement heuristics: (1) linear
placement as a stand-alone placement heuristic; (2) the combination of
matching-based placement and  linear placement; and (3) a random mixture of the
two methods to generate a plurality of placements, followed by performing a
number of random \SWAP gates before calculating the score of each placement.
\mbox{Methods (1) and (2)} each generate a single placement, but method (3) generates
multiple placements that are then compared to each other according to their
placement scores, with the placement that has the best score being selected. We
choose the hyperparmaters  $\delta = 0.1$, $D = 7$, and $\lambda = 40$ in our benchmarking,
performing benchmarks on the test set of circuits provided in~\cite
{li2019tackling}. We select the hyperparameters here with the aim to strike a balance
between performance and runtime for the circuits tested.
We report our results in~\cref
{tab:initial_placement_comparison}.

\begin{table*}[t]
  \centering
  \renewcommand{\arraystretch}{1.1}
  % \resizebox{\columnwidth}{!}{
  \begin{tabular}{r||c|c||c|c|c|c||c|c|c|c}
    \hline
    \multirow{2}{*}{Circuit}
    & No. of & No. of
    & \multicolumn{4}{c||}{Avg. \SWAP Count -- Without Commutation}
    & \multicolumn{4}{c}{Avg. \SWAP Count -- With Commutation}\\
    \cline{4-11}
    & Qubits & Gates & (1) & (2A) & (2B) & (3) & (1) & (2A) & (2B) & (3)\\
    \hline

    4mod5-v1\_22
    & 5 & 21
    & 3 & 3.33 & 2.33 & 2.33
    & \textbf{2}* & 3 & 2.33 & \textbf{2}*\\
    mod5mils\_65
    & 5 & 35
    & 4 & 4 & 4.67 & \textbf{2.33*}
    & 2.67 & 3 & 4.33 & \textbf{2.33*}\\

    alu-v0\_27
    & 5 & 36
    & 3.33 & 5.67 & 6 & 3.33
    & \textbf{2}* & 3.33 & 5.33 & 3.33\\

    decod24-v2\_43
    & 4  & 52
    & 5.67 & 5.67 & 5.67 & \textbf{3.33}*
    & 9.33 & 7.33 & 6 & \textbf{3.33}*\\

    4gt13\_92
    & 5  & 66
    & 8.67 & 7 & 7.67 & 5.67
    & 9.67 & 4.67 & 4.67 & \textbf{4}*\\

    ising\_model\_10
    & 10 & 480
    & \textbf{0}* & 7.33 & 9.67 & 5
    & \textbf{0}* & 5.67 & 8 & 5\\

    ising\_model\_13
    & 13 & 633
    & \textbf{0}* & 12 & 11 & 6.33
    & \textbf{0}* & 10.67 & 11.67 & 6.67\\

    ising\_model\_16
    & 16 & 786
    & 1 & 10.33 & 14 & 7.33
    & \textbf{0}* & 14.67 & 14.33 & 6.33\\

    rd84\_142
    & 15 & 343
    & 49.33 & 59.67 & 52.67 & 43.33
    & \textbf{39.67}* & 47 & 47.67 & 44.33\\

    adr4\_197
    & 13 & 3439
    & 325.67 & 302 & 277.33 & 318
    & 293.67 & 263.67 & \textbf{248.33}* & 248.67\\

    radd\_250
    & 13 & 3213
    & 294.33 & 307 & 311 & 288.33
    & \textbf{227.33}* & 257.33 & 245 & 258\\

    z4\_268
    & 11 & 3073
    & 277.67 & 260.67 & 253.33 & 270
    & \textbf{210.33}* & 243.67 & 237.33 & 225.33\\

    sym6\_145
    & 7  & 3888
    & 324.67 & 314 & 220 & 282.67
    & 231 & 264.67 & \textbf{201}* & 205.67\\

    misex1\_241
    & 15 & 4813
    & 345.33 & 339.67 & 351.33 & 311.33
    & 282 & 275.33 & 326.33 & \textbf{268.67}*\\

    rd73\_252
    & 10 & 5321
    & 430.67 & 480.33 & 486.33 & 412
    & \textbf{375.33}* & 416 & 402 & 396.67\\

    cycle10\_2\_110
    & 12 & 6050
    & 567.67 & 539.33 & 479.33 & 559
    & 489.67 & \textbf{421.67}* & 468.67 & 456.33\\

    square\_root\_7
    & 15 & 7630
    & 558 & \textbf{416}* & 563.33 & 523
    & 431 & 519 & 501 & 420.33\\

    sqn\_258
    & 10 & 10223
    & 937.67 & 838 & 919 & 776
    & 697.67 & \textbf{690.67}* & 747 & 693.33\\

    rd84\_253
    & 12 & 13658
    & 1332.67 & 1202.33 & 1272 & 1170.33
    & 1043.33 & 1011.67 & 1067.67 & \textbf{1010}*\\

    co14\_215
    & 15 & 17936
    & 1861.67 & 1682.33 & 1841 & 1850.67
    & 1665.33 & 1630.67 & \textbf{1438}* & 1639.67\\

    sym9\_193
    & 10 & 34881
    & 3305 & 3272 & 3089.33 & 3249.33
    & 2878.67 & 2782.67 & \textbf{2453}* & 2696.67\\

    9symml\_195
    & 11 & 34881
    & 3113.33 & 2945 & 3153 & 3034
    & 2459 & 2674 & \textbf{2147.33}* & 2817\\ \hline

    \multicolumn{3}{r||}{Number of Lowest \SWAP Counts}
    & {2} & {1} & {0} & {2}
    & {9} & {2} & {5} & {6}\\
    \hline
  \end{tabular}
  \vspace{1mm}
  \caption{\SWAP counts with different placement methods performed on the test
  quantum circuits in~\cite{li2019tackling}.
  Method (1): linear placement,
  method (2A): matching-based placement with greedy matching,
  method (2B): matching-based placement with blossom matching, and
  method (3): multiple placements. The lowest \SWAP count for each circuit is shown in bold text and with an astrisk.}
  \label{tab:initial_placement_comparison}
\end{table*}

As shown in Table~\ref{tab:initial_placement_comparison}, method (1) performs best for
Ising model circuits, independent of whether we apply commutation rules.
This is expected, as the two-qubit gates in the  circuits representing one-dimensional Ising models are \CNOT gates that are applied to successive qubits.
Given the linear structure of Ising model circuits, they are not representative of general benchmarking circuits; therefore,
we exclude them from our discussion and the results
in~\cref{tab:initial_placement_percentage}.

We observe from~\cref
{tab:initial_placement_comparison} that \mbox{method (3)} produces the greatest number
of best results for smaller circuits (i.e., circuits with fewer than 100
gates) among the benchmark tests we run, both with and without the application of commutation rules. However, for
larger circuits (i.e., those having more than 400 gates), method (2B) produces the greatest number
of best results, both with and without applying commutation, by taking advantage
of the matching algorithms and randomness involved in producing the rest of the placements.

\begin{table}[!htbp]
\small
  \renewcommand{\arraystretch}{1.1}
  \begin{tabular}{r||c|c|c|c}
    \hline & (1) & (2A) & (2B) & (3)\\ \hline
    Without Commutation & 44.4\% & 45.5\% & 45.2\% & 24.9\%\\
    With Commutation & 23.5\% & 24.1\% & 25.3\% & 9.1\%\\
    \hline
  \end{tabular}
  \vspace{1mm}
  \caption{Average relative increase in the \SWAP counts for each
   placement method shown in~\cref{tab:initial_placement_comparison}.
  The relative increase percentages are with respect to the baseline, defined as the lowest \SWAP count
  observed between all four methods.
  \mbox{Method (1):} linear placement,
  method (2A): matching-based placement with greedy matching,
  method (2B): matching-based placement with blossom matching, and
  method (3): multiple placements.}
  \label{tab:initial_placement_percentage}
\end{table}
In~\cref
{tab:initial_placement_percentage}, we use the best results as a baseline. We calculate the average percentage increase in the \SWAP count
for each method, as compared to the baseline. On average, method (3) outperforms the other methods, and the use of commutation
rules improves the performance of the transpiler. As a result, we use \mbox{method (3)} and
apply commutation rules for the following benchmark, in which we compare our algorithm's performance against that of other algorithms.

\subsection{Comparison with Other Algorithms}
\label{sec:main-results}

We first compare the results obtained from our transpiler,
called Nuwa, with those produced by \texttt{arct} in~\cite{childs2019circuit}
and~\texttt{tket} in~\cite{sivarajah2020t}. Several families of practically important quantum circuits
have been contributed by~\cite{childs2019circuit, childscircuits} for benchmarking purposes.
We benchmark the three algorithms
on these families, namely, quantum circuits pertaining to quantum signal
processing (QSP)~\cite{low2017optimal}, quantum Fourier \mbox{transformation
(QFT),} and the product formula expansion of exponentials of local Hamiltonians,
which is of interest in Hamiltonian simulation applications~\cite
{childs2021theory}. We also benchmark the three algorithms on randomly generated
circuits. Note that the initial placements for the four algorithms are not identical.

\begin{figure*}[!htbp]
  \centering
  \includegraphics[trim=-0.2cm 0cm -2cm 0cm, clip, width=\linewidth]
  {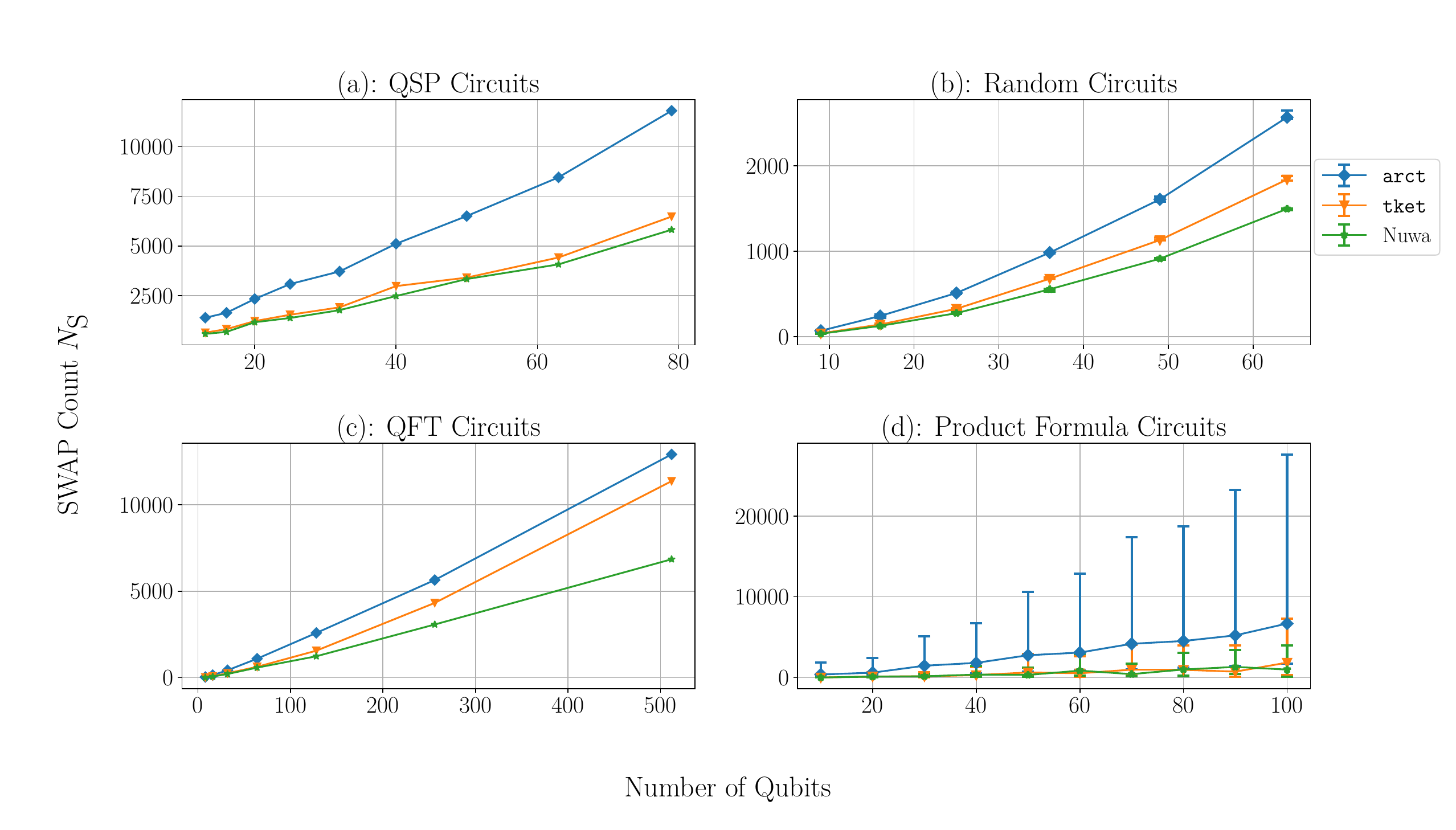}
  \caption{(a) \SWAP counts of the QSP circuits.
  (b) Average \SWAP counts of randomly generated circuits.
  (c) \SWAP counts of QFT circuits.
  (d) Average \SWAP counts of the product formula circuits.
  For (b) and (d), we use the median and interquartile range for each average data point.}
  \label{compare_childs_pytket_nuwa}
\end{figure*}

The results of our comparison are shown in~\cref
{compare_childs_pytket_nuwa}. On average, Nuwa uses 9.0\% fewer \SWAP
gates than \texttt{tket} and 51.8\% fewer \SWAP gates than \texttt{arct} for QSP
circuits. For QFT circuits, the average improvement is 22.4\% and 52.6\%,
respectively. For product formula circuits, Nuwa uses 13.1\% more \SWAP
gates than \texttt{tket} but 83.1\% fewer than \texttt{arct}. For random
circuits, the average improvement is 15.7\% and 45.1\%, respectively.

We next compare Nuwa's results with those of Qiskit
implementation~\cite{hector2019qiskit} of SABRE. We use the same circuits as
used in~\cite{li2019tackling}. Results produced by~\texttt{tket} are also
included, and are shown in \mbox{\cref{tab:other_algo_comparison}.} We note that we
were not able to include SABRE in the study whose results are shown in~\cref{compare_childs_pytket_nuwa}  on test instances of~\cite
{childscircuits}, as the Qiskit implementation of SABRE requires the
number of vertices of the coupling graph to be the same as the number of qubits
in the input quantum circuit.

\begin{table*}[!htbp]
\renewcommand{\arraystretch}{1.1}
    \begin{tabular}{r||c|c||c|c|c|c|c}
      \hline
      \multirow{2}{*}{Circuit}
      & No. of & No. of & \multirow{2}{*}{tket}
      & \multirow{2}{*}{SABRE} & \multirow{2}{*}{arct} & \multicolumn{2}{c}{Nuwa}\\

      \cline{7-8}

      & Qubits & Gates
      & & & & Without Commutation & With Commutation\\

      \hline

      4mod5-v1\_22
      & 5  & 21
      & 4    & 7   & 6  & 2.33    & \textbf{2}*       \\

      mod5mils\_65
      & 5  & 35
      & 3    & 17  & 7   & 2.33    & \textbf{2.33}*    \\

      alu-v0\_27
      & 5  & 36
      & 3    & 12  & 11   & 3.33    & \textbf{3.33}*    \\

      decod24-v2\_43
      & 4  & 52
      & 9    & 20  & 13  & 3.33    & \textbf{3.33}*    \\

      4gt13\_92
      & 5  & 66
      & 9    & 19  & 19   & 5.67    & \textbf{4}*       \\

      ising\_model\_10
      & 10 & 480
      & 9    & 65  & 15   & 5       & \textbf{5}*       \\

      ising\_model\_13
      & 13 & 633
      & 9    & 86  & 29   & \textbf{6.33}*    & 6.67    \\

      ising\_model\_16
      & 16 & 786
      & 16   & 107 & 36   & 7.33    & \textbf{6.33}*    \\

      rd84\_142
      & 15 & 343
      & \textbf{37}* & 78  & 84   & 43.33   & 44.33   \\

      adr4\_197
      & 13 & 3439
      & 440  & 670  & 837  & 318     & \textbf{248.67}*  \\

      radd\_250
      & 13 & 3213
      & 468  & 684 & 828 & 288.33  & \textbf{258}*     \\

      z4\_268
      & 11 & 3073
      & 369  & 713 & 743 & 270     & \textbf{225.33}*  \\

      sym6\_145
      & 7  & 3888
      & 332  & 897  & 792  & 282.67  & \textbf{205.67}*  \\

      misex1\_241
      & 15 & 4813
      & 662  & 1046  & 1132  & 311.33  & \textbf{268.67}*  \\

      rd73\_252
      & 10 & 5321
      & 716  & 1125  & 1208  & 412     & \textbf{396.67}*  \\

      cycle10\_2\_110
      & 12 & 6050
      & 915  & 1178  & 1485  & 559     & \textbf{456.33}*  \\

      square\_root\_7
      & 15 & 7630
      & 1007 & 1240  & 1526  & 523     & \textbf{420.33}*  \\

      sqn\_258
      & 10 & 10223
      & 1100 & 2435  & 2300  & 776     & \textbf{693.33}*  \\

      rd84\_253
      & 12 & 13658
      & 2013 & 2784  & 3060  & 1170.33 & \textbf{1010}*    \\

      co14\_215
      & 15 & 17936
      & 2727 & 3235  & 4860  & 1850.67 & \textbf{1639.67}* \\

      sym9\_193
      & 10 & 34881
      & 5232 & 6211 & 7838  & 3249.33 & \textbf{2696.67}* \\

      9symml\_195
      & 11 & 34881
      & 5232 & 6234 & 7838  & 3034    & \textbf{2817}*   \\

      \hline

      \multicolumn{3}{r||}{Number of Lowest \SWAP Counts}
      & {1} & {0} & {0} & {1}
      & {20}\\
      \hline

    \end{tabular}
  \vspace{1mm}
  \caption{\SWAP counts for the four routing algorithms. The lowest \SWAP count
  for each circuit is shown in bold text and with an astrisk. All tests have been performed on
  the IBM Q20 Tokyo coupling graph, represented in \cref{fig:graph_and_circuit}(b).}
  \label{tab:other_algo_comparison}
\end{table*}

Nuwa outperforms other algorithms by a significant margin, both with and without
applying commutation rules, except for the circuit~\texttt{rd84\_142}.
Without applying commutation rules, Nuwa uses approximately 34\% fewer \SWAP gates than \texttt{tket} on
average, and 66\% fewer than SABRE. With commutation rules applied, the average improvement percentages
are 40\% and 71\%, respectively.

Finally, we note that the benchmarking instances in \cite{li2019tackling} are
extended to a larger set of input cicuits in \cite{qas}. We refer the reader
to \cref{sec:results-tables} for the results of this extended benchmarking
performed using two target architectures.

\section{Conclusion}

The development of quantum circuit compilers and transpilers requires exploration and
incorporation of classical optimization and operations research techniques. Not only are these compilers and transpilers expected to fulfill critical roles in future quantum computing software
stacks, but to serve as important research tools for the production of architecture-aware resource
estimations for quantum algorithms. Transpilers could, in turn, be used to find
improved architectures for quantum devices, and to provide
more-accurate estimates of the number and fidelity of gates required for
practical applications~\mbox{~\cite{sankar2021benchmark, babbush2021focus, gidney2021factor}}.

In this paper, we have introduced a novel transpiler for the placement and
routing of qubits on a target architecture. Our transpiler uses a finite
decision horizon to evaluate two (possibly discounted) rewards: the placement
and edge score functions. It also incorporates a finite lookahead tie-breaking
subroutine. We have benchmarked our transpiler against multiple alternative
solutions and on various test sets. We showed that our algorithm performs significantly
better compared to other algorithms when the circuits contain a large number of gates.

Quantum resources are much more scarce than classical computing resources, which justifies transpilers spending substantial classical computing resources
in the optimization of placement and routing policies. It is, however, important
to have scalable solutions that can be employed on larger quantum circuits or
adapted to fault-tolerant circuit compilation for much larger quantum
algorithms. In addition, it is important for the transpiler to be flexible with
respect to target architectures. This flexibility allows us to use  transpilers as
research tools for quantitative analysis of the overhead of compiling quantum algorithms on various target architectures. Our transpiler provides these critical advantages while improving the state of the art.

\begin{acknowledgements}
This project was funded by 1QBit. P.~R. additionally acknowledges the
financial support of Mike and Ophelia Lazaridis, and Innovation, Science and
Economic Development Canada. All authors thank our editor, Marko Bucyk, for his careful review and editing of the manuscript.
\end{acknowledgements}

\bibliographystyle{ACM-Reference-Format}
\bibliography{main}

\clearpage
\appendix
\section*{Appendix}

\section{Hyperparameter Selection}
\label{sec:param_selection}

We have introduced several hyperparameters in the placement and routing
algorithm, and have studied their effects on the
performance of our transpiler. The hyperparameters we have used are as follows:

\begin{itemize}
\item Discount factor $\delta$:
This discount factor is used in the definitions of the edge scores $\Omega$
and placement scores $\Sigma$. The closer the discount factor is to 1, the more
significant is the contribution of future gates in calculating the scores.

\item Decision horizon $\lambda$:
This hyperparameter determines the number of gates that are to be included in the
calculations of the edge scores $\Omega$ and the placement scores $\Sigma$. A
higher value of $\lambda$ indicates that a greater number of gates will be
included in the score calculations.

\item Lookahead depth $D$:
This hyperparameter determines how many future actions (i.e., \SWAP gate
insertions) the tie-breaking subroutine looks ahead from, within the decision
horizon.
\end{itemize}

To understand how the hyperparameters affect \SWAP count, we generate six
families of circuits: adder, amplitude estimation, Grover's search, phase estimation,
 quantum Fourier transform (QFT), and random. All circuits follow Qiskit's implementation~\cite{hector2019qiskit}.

\begin{figure*}
  \centering
  \includegraphics[trim=0cm 0cm 0cm 0cm, clip, width=\linewidth]
  {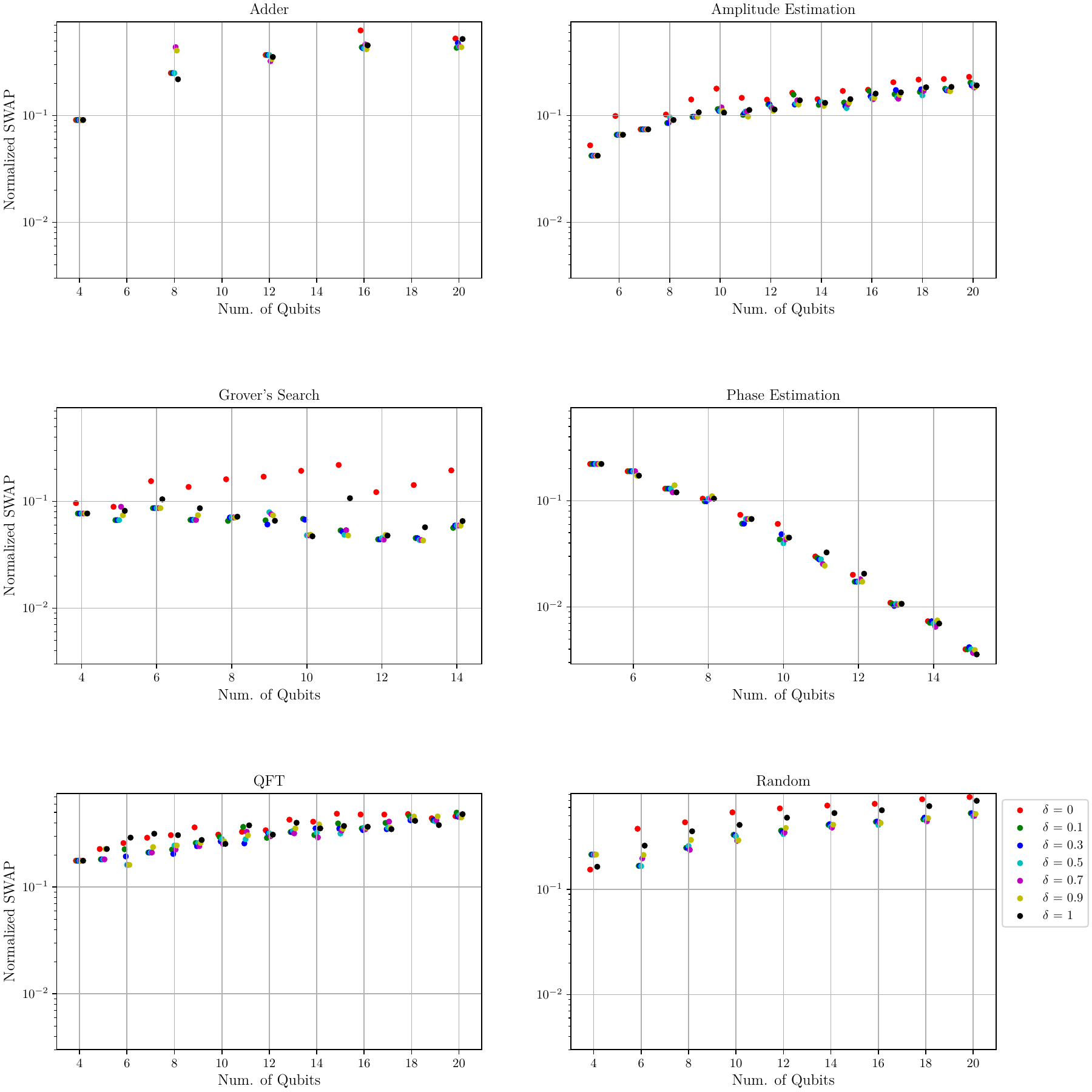}
  \caption{Varying the discount factor $\delta$ with the depth $D=9$ and the decision horizon $\lambda=100$.
  The \SWAP count is normalized by the number of gates in the circuits. We offset the points representing the normalized \SWAP values in the $x$ direction for visual clarity.}
  \label{varying_discount_d9_lambda100}
\end{figure*}

We first vary the value of $\delta$ and fix the decision horizon at
$\lambda=100$ and lookahead depth at $D=9$. From~\cref{varying_discount_d9_lambda100},
higher values of $\delta$ in general result in lower \SWAP counts.
We refer to the $\delta$ values of 0.7, 0.9, and 1 as \emph{large} $\delta$ values.
For adder circuits with 4, 8, 12, and 16 qubits, we attain the lowest \SWAP counts
using large $\delta$, while for the adder circuit with 20 qubits, a value of $\delta=0.1$ results
in the lowest \SWAP count, followed closely by $\delta=0.9$. In other words, using a
large $\delta$ value results in the lowest \SWAP count in $80\%$ of the test cases for adder circuits.
For the amplitude estimation, Grover's search, phase estimation, QFT, and random circuits, using a large $\delta$ value,
we attain a lowest \SWAP count of $81.25\%$, $63.63\%$, $80\%$, $70.59\%$, and $55.56\%$ respectively.

\begin{figure*}
  \centering
  \includegraphics[trim=0cm 0cm 0cm 0cm, clip, width=\linewidth]
  {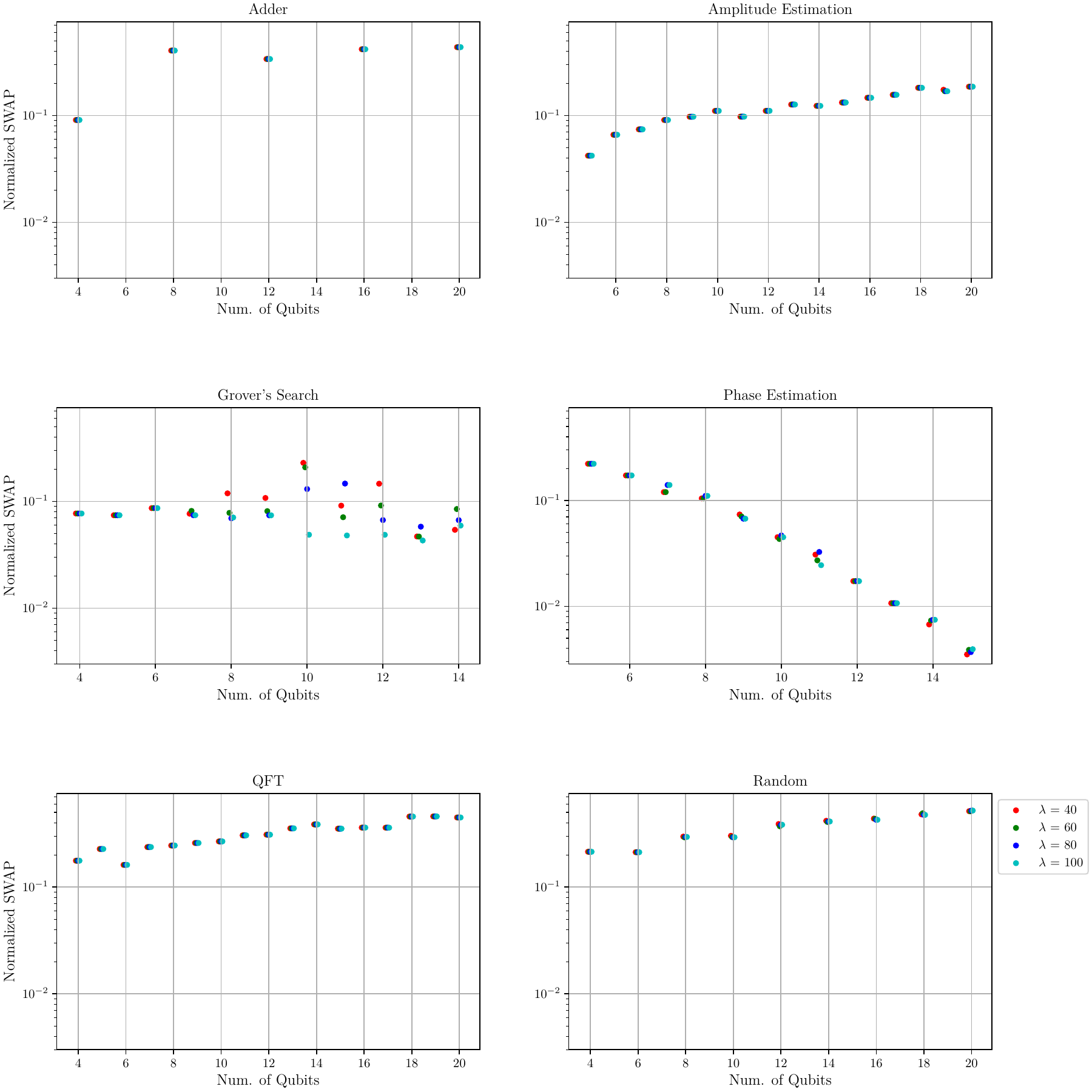}
  \caption{Varying the decision horizon $\lambda$ with the depth $D=9$ and the discount factor $\delta=0.9$.
  The \SWAP count is normalized by the number of gates in the circuits. We offset the points representing the normalized \SWAP values in the $x$ direction for visual clarity.}
  \label{vary_lambda}
\end{figure*}

\Cref{vary_lambda} shows that the value of $\lambda$ does not affect the \SWAP count in the case of
the adder, amplitude estimation,
QFT, and random circuits. As for the Grover's search and phase estimation circuits, we observe that a higher value of
$\lambda$ usually results in a lower \SWAP count, except for the Grover's search circuit with 14 qubits, and the
phase estimation circuits with 7, 8, 10, 14, and 15 qubits. However, even for circuits that do not result in the lowest
\SWAP count using high values of $\lambda$, we observe that the difference in \SWAP count is not significant.
Therefore, we conclude that having a higher value of $\lambda$ is in general beneficial.

\begin{figure*}
  \centering
  \includegraphics[trim=0cm 0cm 0cm 0cm, clip, width=\linewidth]
  {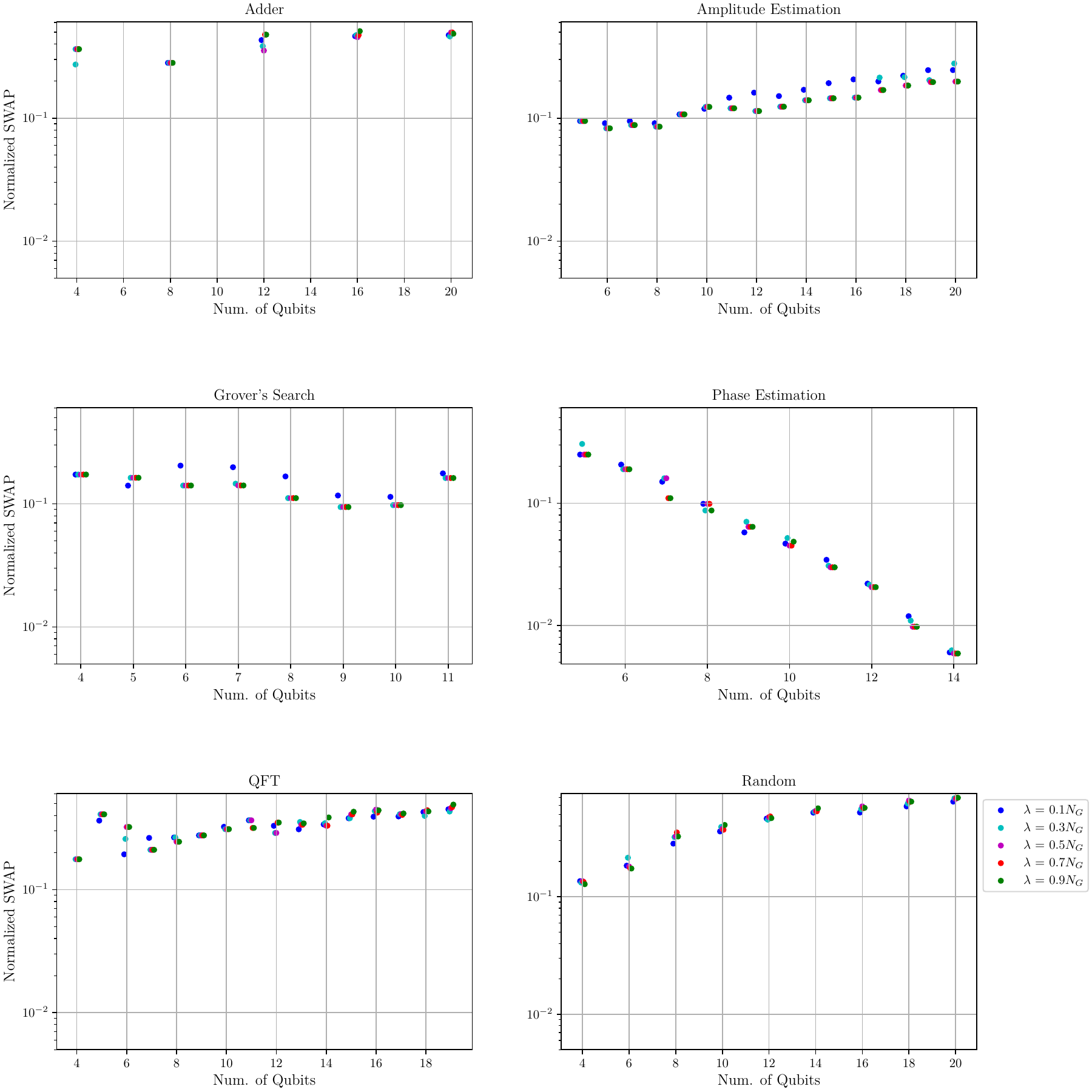}
  \caption{Varying the decisison horizon $\lambda$ with the discount factor $\delta=0.9$
  and the lookahead depth $D=0.5\lambda$. The total number of gates in a test circuit is denoted by $N_\text{G}$. The \SWAP count is normalized by the number of gates in the circuits. We offset the points representing the normalized \SWAP values in the $x$ direction for visual clarity.}
  \label{vary_lambda_D}
\end{figure*}

Having understood that $\lambda$ does not on its own affect \SWAP counts significantly, we design another
experiment where we fix the ratio between $\lambda$ and $D$. We set $\lambda$ to a certain percentage
of the number of gates in the circuits and fix the value of $D$ to be half that of $\lambda$.
The result is shown in~\cref{vary_lambda_D}.
We observe that for the amplitude estimation, Grover's search, and phase estimation circuits,
higher values of $\lambda$ and $D$ result in the majority of cases where the \SWAP count is lowest, while the results
for the adder, QFT, and random circuits are inconsistent. It is worth noting that the results for the
amplitude estimation and Grover's search circuits overlap to a great extent, indicating that the circuits are not
sensitive to the value of either $\lambda$ or $D$; in other words, once the values of $\lambda$ and $D$ exceed
a certain threshold, the \SWAP count will not decrease further even if larger values of $\lambda$ and $D$ are used. We suspect that this is due to the structure of the amplitude estimation and Grover's search circuits,
where a lot of two-qubit gates are situated between single-qubit gates. When the two-qubit gates
do not commute with the single-qubit gates, they need to be executed in sequential order. Thus, the effects
of $\lambda$ and $D$ become negligible.

\clearpage
\section{Detailed Benchmarking Results}
\label{sec:results-tables}

\begin{figure*}[!htbp]
  \centering
  \includegraphics[trim=-0.2cm 0cm -2cm 0cm, clip, width=\linewidth]
  {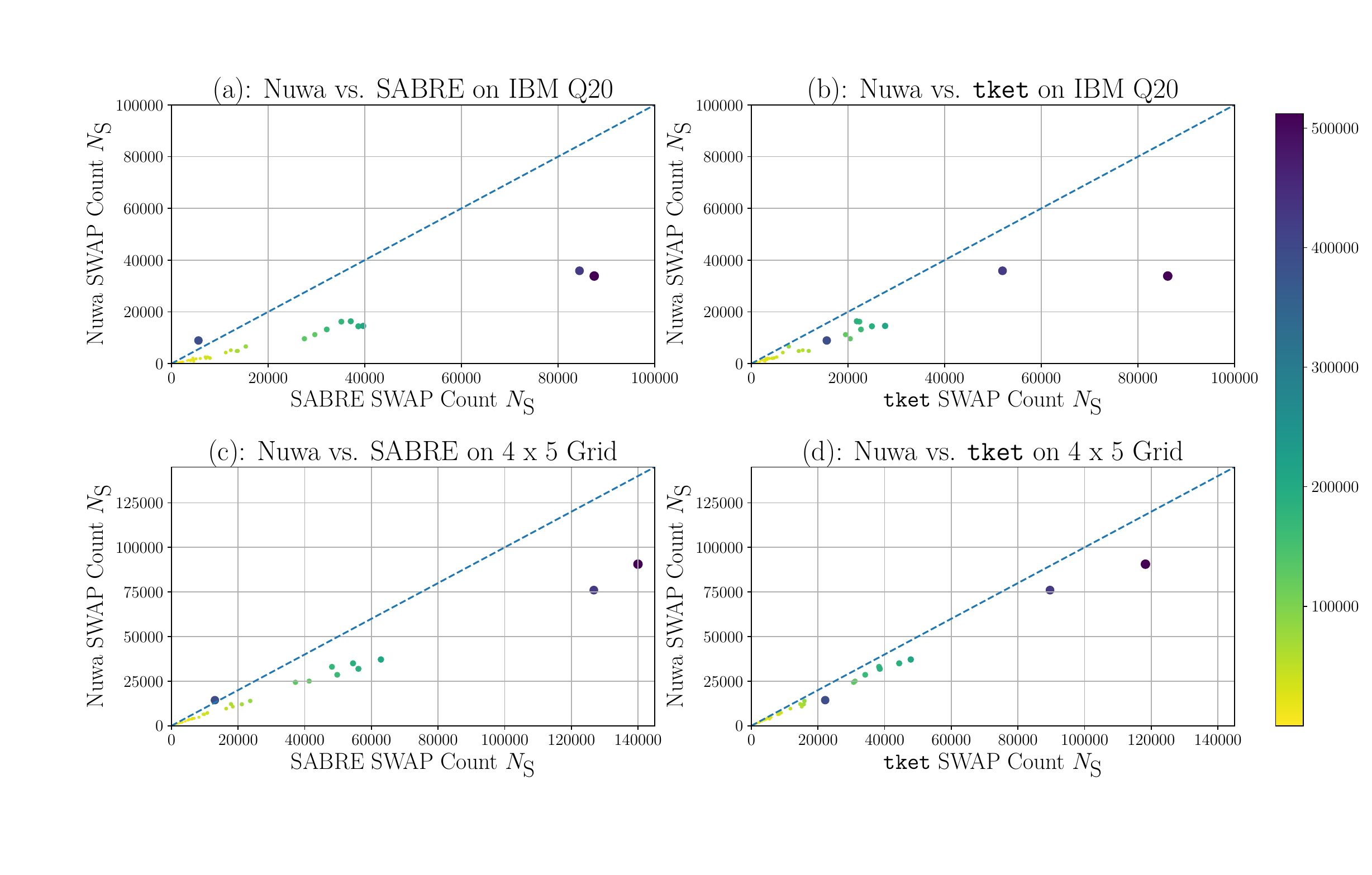}
  \caption{A plot corresponding to \cref{tab:large_benchmark}.
  (a) \SWAP counts of Nuwa vs. SABRE on the IBM Q20.
  (b) \SWAP counts of Nuwa vs. \texttt{tket} on the IBM Q20.
  (c) \SWAP counts of Nuwa vs. SABRE on a 4 x 5 grid.
  (d) \SWAP counts of Nuwa vs. \texttt{tket} on a 4 x 5 grid.
  The colours and sizes of the dots correspond to the number of gates in the circuit.}
  \label{fig:compare_sabre_pytket_nuwa_plot}
\end{figure*}

\cref{tab:large_benchmark} shows the
detailed benchmarking results for \texttt{tket}, SABRE, and Nuwa
transpiling input circuits onto the coupling graphs of two target architechtures: the IBM Q20
Tokyo chip and a 4 x 5 grid. In order to study the routing
algorithms in isolation, we use a na\"ive fixed placement (i.e.,
matching the vertex and qubit numbers) across all experiments. We provide a visualization
of these results in \cref{fig:compare_sabre_pytket_nuwa_plot}. The benchmarking
circuit instances may be found in \cite{qas}.

\begin{longtable*}{r||c|c||c|c|c||c|c|c}
\hline
Input Circuit & No. of & No. of &
\multicolumn{3}{c||}{IBM Q20 Tokyo Graph} &
\multicolumn{3}{c}{4 $\times$ 5 Grid} \\
\cline{4-9}
Name& Qubits & Gates & \texttt{tket} & SABRE & Nuwa & \texttt{tket} & SABRE & Nuwa \\
\hline
xor5\_254 & 6 & 7 & \textbf{1}* & 5 & 3 & \textbf{2}* & 5 & 3 \\
graycode6\_47 & 6 & 5 & \textbf{1}* & 2 & 6 & \textbf{2}* & 3 & 5 \\
ex1\_226 & 6 & 7 & \textbf{1}* & 2 & 3 & \textbf{2}* & 4 & 3 \\
4gt11\_84 & 4 & 18 & 3 & 7 & \textbf{2}* & 3 & 7 & \textbf{2}* \\
4mod5-v0\_20 & 5 & 20 & \textbf{3}* & 4 & \textbf{3}* & \textbf{3}* & 4 & 5 \\
ex-1\_166 & 3 & 19 & \textbf{3}* & 6 & \textbf{3}* & \textbf{3}* & 6 & \textbf{3}* \\
4mod5-v1\_22 & 5 & 21 & 4 & 7 & \textbf{1}* & 4 & 7 & \textbf{3}* \\
mod5d1\_63 & 5 & 22 & \textbf{1}* & 12 & 4 & 5 & 13 & \textbf{4}* \\
ham3\_102 & 3 & 20 & 3 & 6 & \textbf{2}* & 3 & 6 & \textbf{2}* \\
4gt11\_83 & 5 & 23 & \textbf{4}* & 10 & \textbf{4}* & \textbf{4}* & 7 & \textbf{4}* \\
4gt11\_82 & 5 & 27 & \textbf{5}* & 11 & \textbf{5}* & 6 & 10 & \textbf{4}* \\
rd32-v0\_66 & 4 & 34 & 6 & 14 & \textbf{1}* & \textbf{6}* & 14 & \textbf{6}* \\
alu-v0\_27 & 5 & 36 & \textbf{3}* & 13 & 4 & 7 & 12 & \textbf{6}* \\
4mod5-v1\_24 & 5 & 36 & 8 & 12 & \textbf{6}* & 8 & 14 & \textbf{7}* \\
4mod5-v0\_19 & 5 & 35 & \textbf{3}* & 16 & 5 & \textbf{7}* & 16 & 10 \\
mod5mils\_65 & 5 & 35 & \textbf{3}* & 17 & 5 & \textbf{6}* & 17 & 10 \\
rd32-v1\_68 & 4 & 36 & 6 & 14 & \textbf{1}* & \textbf{6}* & 14 & \textbf{6}* \\
alu-v1\_28 & 5 & 37 & \textbf{4}* & 12 & 5 & 8 & 12 & \textbf{7}* \\
alu-v2\_33 & 5 & 37 & \textbf{3}* & 10 & \textbf{3}* & \textbf{6}* & 10 & 8 \\
alu-v4\_37 & 5 & 37 & \textbf{3}* & 14 & 4 & 7 & 17 & \textbf{6}* \\
alu-v3\_35 & 5 & 37 & \textbf{3}* & 10 & 4 & 7 & 12 & \textbf{6}* \\
3\_17\_13 & 3 & 36 & 6 & 13 & \textbf{1}* & 6 & 11 & \textbf{5}* \\
alu-v1\_29 & 5 & 37 & \textbf{4}* & 10 & \textbf{4}* & 8 & 10 & \textbf{6}* \\
miller\_11 & 3 & 50 & 9 & 19 & \textbf{0}* & \textbf{9}* & 19 & \textbf{9}* \\
alu-v3\_34 & 5 & 52 & 6 & 20 & \textbf{3}* & \textbf{10}* & 25 & 12 \\
decod24-v2\_43 & 4 & 52 & 9 & 20 & \textbf{2}* & 9 & 20 & \textbf{8}* \\
decod24-v0\_38 & 4 & 51 & \textbf{6}* & 17 & 7 & 9 & 18 & \textbf{8}* \\
mod5d2\_64 & 5 & 53 & \textbf{6}* & 26 & 8 & \textbf{11}* & 25 & 12 \\
4gt13\_92 & 5 & 66 & \textbf{9}* & 21 & \textbf{9}* & 12 & 29 & \textbf{11}* \\
4gt13-v1\_93 & 5 & 68 & \textbf{6}* & 24 & 7 & 12 & 27 & \textbf{9}* \\
4mod5-v0\_18 & 5 & 69 & \textbf{4}* & 31 & 9 & \textbf{12}* & 31 & 14 \\
decod24-bdd\_294 & 6 & 73 & \textbf{9}* & 20 & 13 & \textbf{13}* & 23 & 14 \\
one-two-...v2\_100 & 5 & 69 & 7 & 22 & \textbf{4}* & 14 & 23 & \textbf{11}* \\
one-two-...v3\_101 & 5 & 70 & \textbf{7}* & 21 & 8 & \textbf{13}* & 21 & 14 \\
4mod5-v1\_23 & 5 & 69 & \textbf{4}* & 35 & 9 & 14 & 33 & \textbf{12}* \\
4mod5-bdd\_287 & 7 & 70 & \textbf{7}* & 13 & 8 & \textbf{14}* & 24 & 15 \\
rd32\_270 & 5 & 84 & \textbf{6}* & 34 & 11 & \textbf{16}* & 34 & \textbf{16}* \\
4gt5\_75 & 5 & 83 & \textbf{9}* & 34 & 11 & 16 & 34 & \textbf{12}* \\
alu-bdd\_288 & 7 & 84 & 15 & 11 & \textbf{9}* & \textbf{16}* & 17 & 17 \\
alu-v0\_26 & 5 & 84 & 10 & 34 & \textbf{9}* & 16 & 34 & \textbf{14}* \\
decod24-v1\_41 & 5 & 85 & 14 & 26 & \textbf{6}* & 17 & 34 & \textbf{16}* \\
rd53\_138 & 8 & 132 & \textbf{11}* & 23 & 12 & \textbf{25}* & 38 & 31 \\
4gt5\_76 & 5 & 91 & 14 & 28 & \textbf{12}* & \textbf{17}* & 40 & \textbf{17}* \\
4gt13\_91 & 5 & 103 & 13 & 39 & \textbf{12}* & \textbf{19}* & 38 & 21 \\
cnt3-5\_179 & 16 & 175 & 31 & 26 & \textbf{24}* & 52 & \textbf{38}* & 40 \\
qft\_10 & 10 & 200 & \textbf{16}* & 57 & 22 & \textbf{24}* & 60 & 25 \\
4gt13\_90 & 5 & 107 & 15 & 43 & \textbf{12}* & \textbf{20}* & 40 & 21 \\
alu-v4\_36 & 5 & 115 & \textbf{13}* & 41 & \textbf{13}* & 20 & 43 & \textbf{18}* \\
mini\_alu\_305 & 10 & 173 & 27 & 41 & \textbf{22}* & 39 & 46 & \textbf{34}* \\
ising\_model\_10 & 10 & 480 & 9 & 66 & \textbf{4}* & 15 & 30 & \textbf{11}* \\
ising\_model\_16 & 16 & 786 & \textbf{11}* & 76 & 17 & 46 & 75 & \textbf{16}* \\
ising\_model\_13 & 13 & 633 & \textbf{9}* & 43 & 15 & \textbf{26}* & 82 & 27 \\
4gt5\_77 & 5 & 131 & 18 & 51 & \textbf{13}* & \textbf{22}* & 51 & 24 \\
sys6-v0\_111 & 10 & 215 & 22 & 44 & \textbf{21}* & 50 & 54 & \textbf{45}* \\
one-two-...v1\_99 & 5 & 132 & 15 & 59 & \textbf{11}* & \textbf{27}* & 58 & 28 \\
one-two-...v0\_98 & 5 & 146 & 16 & 47 & \textbf{11}* & 25 & 64 & \textbf{21}* \\
decod24-v3\_45 & 5 & 150 & 19 & 42 & \textbf{15}* & \textbf{25}* & 57 & 27 \\
4gt10-v1\_81 & 5 & 148 & 24 & 40 & \textbf{11}* & \textbf{27}* & 62 & 29 \\
aj-e11\_165 & 5 & 151 & 27 & 45 & \textbf{14}* & 28 & 55 & \textbf{26}* \\
4mod7-v0\_94 & 5 & 162 & \textbf{17}* & 45 & 19 & 30 & 63 & \textbf{27}* \\
alu-v2\_32 & 5 & 163 & 23 & 57 & \textbf{14}* & \textbf{29}* & 64 & 30 \\
rd73\_140 & 10 & 230 & \textbf{19}* & 57 & 24 & 51 & 57 & \textbf{49}* \\
4mod7-v1\_96 & 5 & 164 & 24 & 45 & \textbf{10}* & 31 & 64 & \textbf{28}* \\
4gt4-v0\_80 & 6 & 179 & \textbf{11}* & 36 & 17 & 34 & 42 & \textbf{33}* \\
mod10\_176 & 5 & 178 & 26 & 58 & \textbf{18}* & 33 & 75 & \textbf{28}* \\
0410184\_169 & 14 & 211 & \textbf{33}* & 50 & 36 & 58 & \textbf{55}* & 56 \\
qft\_16 & 16 & 512 & 55 & 137 & \textbf{40}* & 79 & 142 & \textbf{55}* \\
4gt12-v0\_88 & 6 & 194 & 31 & 41 & \textbf{22}* & 45 & 46 & \textbf{33}* \\
rd84\_142 & 15 & 343 & 37 & 57 & \textbf{36}* & 74 & \textbf{62}* & 72 \\
rd53\_311 & 13 & 275 & 51 & 56 & \textbf{26}* & 64 & 94 & \textbf{62}* \\
4\_49\_16 & 5 & 217 & 33 & 76 & \textbf{21}* & 44 & 87 & \textbf{36}* \\
sym9\_146 & 12 & 328 & \textbf{30}* & 59 & 31 & 70 & 100 & \textbf{60}* \\
4gt12-v1\_89 & 6 & 228 & 26 & 76 & \textbf{14}* & 41 & 64 & \textbf{40}* \\
4gt12-v0\_87 & 6 & 247 & \textbf{21}* & 31 & 28 & 45 & 73 & \textbf{41}* \\
4gt4-v0\_79 & 6 & 231 & \textbf{15}* & 35 & 30 & 43 & 67 & \textbf{39}* \\
hwb4\_49 & 5 & 233 & \textbf{22}* & 94 & 24 & 41 & 94 & \textbf{37}* \\
sym6\_316 & 14 & 270 & 43 & 62 & \textbf{41}* & \textbf{59}* & 79 & 67 \\
4gt12-v0\_86 & 6 & 251 & \textbf{22}* & 31 & 30 & 46 & 76 & \textbf{41}* \\
4gt4-v0\_72 & 6 & 258 & 28 & 45 & \textbf{27}* & 53 & 99 & \textbf{44}* \\
4gt4-v0\_78 & 6 & 235 & \textbf{17}* & 36 & 32 & 44 & 68 & \textbf{40}* \\
mod10\_171 & 5 & 244 & 33 & 81 & \textbf{25}* & 44 & 89 & \textbf{43}* \\
4gt4-v1\_74 & 6 & 273 & 37 & 61 & \textbf{27}* & 64 & 87 & \textbf{45}* \\
rd53\_135 & 7 & 296 & \textbf{23}* & 76 & \textbf{23}* & 59 & 88 & \textbf{56}* \\
mini-alu\_167 & 5 & 288 & 38 & 96 & \textbf{35}* & 53 & 115 & \textbf{49}* \\
one-two-...v0\_97 & 5 & 290 & 41 & 122 & \textbf{36}* & 56 & 121 & \textbf{48}* \\
ham7\_104 & 7 & 320 & 43 & 61 & \textbf{17}* & 62 & 82 & \textbf{58}* \\
decod24-enable... & 6 & 338 & 35 & 75 & \textbf{20}* & 62 & 124 & \textbf{58}* \\
mod8-10\_178 & 6 & 342 & 29 & 71 & \textbf{23}* & 78 & 105 & \textbf{59}* \\
cnt3-5\_180 & 16 & 485 & 83 & 127 & \textbf{32}* & 107 & 130 & \textbf{104}* \\
ex3\_229 & 6 & 403 & 52 & 81 & \textbf{39}* & 92 & 115 & \textbf{64}* \\
4gt4-v0\_73 & 6 & 395 & 43 & 89 & \textbf{25}* & 79 & 145 & \textbf{72}* \\
mod8-10\_177 & 6 & 440 & 44 & 97 & \textbf{35}* & 86 & 115 & \textbf{72}* \\
C17\_204 & 7 & 467 & 70 & 134 & \textbf{25}* & \textbf{89}* & 167 & 91 \\
alu-v2\_31 & 5 & 451 & \textbf{42}* & 141 & \textbf{42}* & 90 & 196 & \textbf{76}* \\
rd53\_131 & 7 & 469 & 46 & 94 & \textbf{33}* & 94 & 147 & \textbf{88}* \\
alu-v2\_30 & 6 & 504 & 59 & 104 & \textbf{39}* & 96 & 181 & \textbf{93}* \\
mod5adder\_127 & 6 & 555 & 83 & 102 & \textbf{49}* & 121 & 184 & \textbf{96}* \\
rd53\_133 & 7 & 580 & 108 & 195 & \textbf{26}* & 121 & 156 & \textbf{98}* \\
cm82a\_208 & 8 & 650 & 69 & 160 & \textbf{29}* & \textbf{120}* & 180 & 125 \\
majority\_239 & 7 & 612 & 75 & 176 & \textbf{42}* & 121 & 192 & \textbf{105}* \\
ex2\_227 & 7 & 631 & 89 & 194 & \textbf{31}* & 119 & 220 & \textbf{101}* \\
sf\_276 & 6 & 778 & 76 & 127 & \textbf{54}* & 147 & 262 & \textbf{127}* \\
sf\_274 & 6 & 781 & 84 & 209 & \textbf{76}* & 172 & 201 & \textbf{137}* \\
con1\_216 & 9 & 954 & 178 & 227 & \textbf{40}* & 193 & 307 & \textbf{170}* \\
wim\_266 & 11 & 986 & 125 & 224 & \textbf{78}* & 227 & 307 & \textbf{184}* \\
rd53\_130 & 7 & 1043 & 106 & 219 & \textbf{72}* & 239 & 290 & \textbf{184}* \\
f2\_232 & 8 & 1206 & 194 & 343 & \textbf{73}* & 252 & 343 & \textbf{217}* \\
cm152a\_212 & 12 & 1221 & 202 & 260 & \textbf{79}* & 273 & 354 & \textbf{213}* \\
rd53\_251 & 8 & 1291 & 172 & 205 & \textbf{100}* & 278 & 406 & \textbf{233}* \\
hwb5\_53 & 6 & 1336 & 153 & 328 & \textbf{110}* & 254 & 397 & \textbf{245}* \\
cm42a\_207 & 14 & 1776 & 236 & 412 & \textbf{92}* & 374 & 489 & \textbf{305}* \\
pm1\_249 & 14 & 1776 & 236 & 285 & \textbf{92}* & 374 & 558 & \textbf{305}* \\
dc1\_220 & 11 & 1914 & 278 & 309 & \textbf{92}* & 396 & 610 & \textbf{338}* \\
squar5\_261 & 13 & 1993 & 339 & 392 & \textbf{128}* & 445 & 546 & \textbf{355}* \\
z4\_268 & 11 & 3073 & 369 & 668 & \textbf{216}* & 750 & 874 & \textbf{537}* \\
sqrt8\_260 & 12 & 3009 & 476 & 552 & \textbf{252}* & 725 & 881 & \textbf{575}* \\
radd\_250 & 13 & 3213 & 468 & 635 & \textbf{205}* & 708 & 939 & \textbf{608}* \\
adr4\_197 & 13 & 3439 & 495 & 641 & \textbf{224}* & 719 & 936 & \textbf{638}* \\
sym6\_145 & 7 & 3888 & 332 & 1043 & \textbf{214}* & 759 & 1107 & \textbf{680}* \\
misex1\_241 & 15 & 4813 & 589 & 938 & \textbf{253}* & 993 & 1407 & \textbf{909}* \\
rd73\_252 & 10 & 5321 & 690 & 1043 & \textbf{389}* & 1118 & 1664 & \textbf{976}* \\
cycle10\_2\_110 & 12 & 6050 & 910 & 1276 & \textbf{423}* & 1354 & 1624 & \textbf{1089}* \\
hwb6\_56 & 7 & 6723 & 956 & 1438 & \textbf{409}* & 1321 & 1998 & \textbf{1184}* \\
square\_root\_7 & 15 & 7630 & 1001 & 1118 & \textbf{504}* & 1640 & 2156 & \textbf{1301}* \\
ham15\_107 & 15 & 8763 & 1163 & 1636 & \textbf{589}* & 1909 & 2408 & \textbf{1615}* \\
dc2\_222 & 15 & 9462 & 1200 & 2086 & \textbf{714}* & 2110 & 2599 & \textbf{1752}* \\
sqn\_258 & 10 & 10223 & 1100 & 2416 & \textbf{709}* & 2352 & 3212 & \textbf{1870}* \\
inc\_237 & 16 & 10619 & 1232 & 1904 & \textbf{600}* & 2351 & 2964 & \textbf{1888}* \\
cm85a\_209 & 14 & 11414 & 1720 & 2173 & \textbf{699}* & 2553 & 3315 & \textbf{2076}* \\
rd84\_253 & 12 & 13658 & 1914 & 3265 & \textbf{1188}* & 3130 & 4132 & \textbf{2537}* \\
co14\_215 & 15 & 17936 & 2583 & 3922 & \textbf{1054}* & 4364 & 5220 & \textbf{3577}* \\
root\_255 & 13 & 17159 & 2547 & 3535 & \textbf{1328}* & 3824 & 4915 & \textbf{3152}* \\
mlp4\_245 & 16 & 18852 & 2168 & 4073 & \textbf{1463}* & 4223 & 5396 & \textbf{3572}* \\
urf2\_277 & 8 & 20112 & 3321 & 4434 & \textbf{2191}* & 5543 & 6469 & \textbf{4091}* \\
sym9\_148 & 10 & 21504 & 2852 & 4596 & \textbf{886}* & 5473 & 6060 & \textbf{3706}* \\
life\_238 & 11 & 22445 & 3266 & 4570 & \textbf{1551}* & 5287 & 6272 & \textbf{4157}* \\
hwb7\_59 & 8 & 24379 & 2956 & 5076 & \textbf{1854}* & 4768 & 6791 & \textbf{4196}* \\
max46\_240 & 10 & 27126 & 3697 & 5941 & \textbf{1984}* & 5894 & 8226 & \textbf{4796}* \\
clip\_206 & 14 & 33827 & 4505 & 7147 & \textbf{2063}* & 7958 & 9833 & \textbf{6319}* \\
9symml\_195 & 11 & 34881 & 5232 & 7504 & \textbf{2537}* & 8268 & 9545 & \textbf{6441}* \\
sym9\_193 & 11 & 34881 & 5232 & 6995 & \textbf{2537}* & 8268 & 9803 & \textbf{6441}* \\
sao2\_257 & 14 & 38577 & 4312 & 7966 & \textbf{2083}* & 8938 & 10798 & \textbf{7310}* \\
dist\_223 & 13 & 38046 & 4707 & 7939 & \textbf{2182}* & 8582 & 10704 & \textbf{7075}* \\
urf5\_280 & 9 & 49829 & 6520 & 11229 & \textbf{4277}* & 11720 & 16416 & \textbf{9596}* \\
urf1\_278 & 9 & 54766 & 10646 & 12291 & \textbf{5187}* & 15126 & 18407 & \textbf{10727}* \\
sym10\_262 & 12 & 64283 & 9810 & 13513 & \textbf{4872}* & 14673 & 17856 & \textbf{12177}* \\
hwb8\_113 & 9 & 69380 & 11874 & 13697 & \textbf{4904}* & 15758 & 21123 & \textbf{12062}* \\
urf2\_152 & 8 & 80480 & 7748 & 15376 & \textbf{6612}* & 15923 & 23620 & \textbf{13928}* \\
urf3\_279 & 10 & 125362 & 19498 & 29663 & \textbf{11219}* & 31126 & 41320 & \textbf{25021}* \\
plus63mod...163 & 13 & 128744 & 20470 & 27511 & \textbf{9618}* & 30692 & 37207 & \textbf{24375}* \\
urf5\_158 & 9 & 164416 & 22682 & 32133 & \textbf{13220}* & 34169 & 49734 & \textbf{28583}* \\
urf6\_160 & 15 & 171840 & 22359 & 35154 & \textbf{16218}* & 38318 & 48154 & \textbf{33050}* \\
urf1\_149 & 9 & 184864 & 21828 & 37104 & \textbf{16367}* & 38558 & 56099 & \textbf{31928}* \\
plus63mod...164 & 14 & 187112 & 24956 & 38682 & \textbf{14438}* & 44366 & 54477 & \textbf{35014}* \\
hwb9\_119 & 10 & 207775 & 27687 & 39659 & \textbf{14569}* & 47846 & 62845 & \textbf{37163}* \\
urf3\_155 & 10 & 423488 & 51978 & 84436 & \textbf{35888}* & 89626 & 126737 & \textbf{76050}* \\
ground\_...10 & 13 & 390180 & 15612 & \textbf{5586}* & 8947 & 22169 & \textbf{13030}* & 14378 \\
urf4\_187 & 11 & 512064 & 86174 & 87487 & \textbf{33856}* & 118263 & 139987 & \textbf{90583}* \\
\hline
\multicolumn{3}{r||}{Number of Lowest \SWAP Counts} &
44 & 1 & \textbf{123}* & 34 & 4 & \textbf{127}* \\
\hline
\caption{\SWAP counts for \texttt{tket}, SABRE, and Nuwa. The lowest \SWAP count achieved for each circuit is shown in bold text and with an
astrisk. Columns 4 to 6 show the results for the IBM Q20 Tokyo
coupling graph represented in \cref{fig:graph_and_circuit}(b) and
columns 7 to 9 show tests performed on a 4 $\times$ 5 target architecture.}
\label{tab:large_benchmark}
\end{longtable*}

\end{document}